# Design principles for metal-organic receptors targeting optical recognition of Pd(II) in environmental matrices

Sudhanshu Naithani,[a] Pramod Kumar,[b] Ritesh Dubey,[a] Franck Thetiot,*[,c] Samar Layek,*[,d] Tapas Goswami,*[,a] Sushil Kumar*[,a]

A precise detection of palladium (Pd) ions is a critical challenge with significant socio-economic implications across various industrial and chemical sectors. Due to its widespread use and poor biodegradability, $Pd^{2+}$ accumulates in environmental ecosystems, posing severe risks to both the environment and living organisms. Consequently, there is a strong demand for selective, sensitive, and user-friendly detection methods. Among emerging strategies, optical detection techniques (both luminescent and colorimetric) using metal-based receptors have gained considerable attention. These sensors offer distinct advantages over traditional organic probes, including large Stokes shifts, long emission lifetimes, exceptional photostability, enhanced water solubility, recyclability, and remarkable chemical versatility. These attributes make them highly suitable for diverse applications in sensing and bioanalytical fields. This review provides a comprehensive overview of recent advancements in luminescent and colorimetric metal-based probes, including metal complexes and metal-organic frameworks (MOFs), for the selective detection of $Pd^{2+}$. It discusses key design strategies, critical performance factors, and future prospects, offering valuable insights for researchers working on next-generation sensing platform.

## 1. Introduction

Palladium (Pd), a platinum group element, is widely utilized in various chemical, industrial, environmental, and biological applications due to its exceptional chemical and physical properties.[1–4] Over the preceding years, Pd-based catalysts have demonstrated remarkable efficiency and gained prominence, particularly in organic chemistry across both small as well as large-scale industrial operations.[5–8] The widespread use of Pd is driven by its remarkable ability to form C-C and C-O bonds *via* key organic reactions such as Suzuki-Miyaura, Heck, Negishi, Sonogashira, and Stille coupling reactions.[9–11] These reactions are essential for a range of industrial and environmental applications, including the production of pharmaceuticals, polymers, fine chemicals, agrochemicals, to name but a few.[9–14] The recurring overuse of this metal has parallelly led to the release of residual contaminants into different environmental matrices, particularly soils and water bodies, with serious environmental consequences.[15,16] Although palladium is not a naturally occurring bio-metal, real-time analysis and imaging of this metal have gained raising attention due to its potential health risks and environmental impact.[17–19] Indeed, this heavy metal exhibits a strong binding affinity for heteroatoms such as S, Se, N, and O, meaning that even a small intake of Pd-based ions in living organisms may lead to substantial biological dysfunctions.[20–23] Hence, a dilemma has emerged where the socio-economic benefits of using Pd are increasingly overshadowed by the negative impacts of its overuse/mishandling. This situation underscores the urgent need to develop selective and efficient methods for both qualitative and quantitative Pd detection and diagnosis, prior to implementing appropriate treatment solutions.[24,25] However, this effort is complicated by the chemical similarities between palladium and other elements within the same group, especially platinum, in several targeted media.

In terms of detection, several techniques such as inductively coupled plasma optical emission spectrometry (ICP-OES), atomic absorption spectrophotometry (AAS), high performance liquid chromatography (HPLC) and electrochemical analyses, are largely employed to quantify metal ion contaminants.[26–32] These techniques are typically favored for their remarkable analytical sensitivity with the ability to selectively detect metals at low concentrations (down to the ppb level). But the sophisticated instrumentation and related labour-intensive sample preparation procedures significantly restrict their practicality for many real-time and cost-effective applications.[33–36] As pertinent alternatives, the optical sensing methods have consequently gained much attention from the researchers. Indeed, the optical sensors offer some unique advantages, including high sensitivity/selectivity, remarkably low detection limits (LoD), fast analyte response, operational simplicity, cost-effectiveness, miniaturization potential, as well as compatibility with real-time and on-site monitoring.[37–39]

Optical sensors can be typically sub-divided into two main categories: (i) luminescent sensors, which exhibit distinct luminescence response in the presence of a target guest analyte (e.g., $Pd^{2+}$ in this study), making them invaluable for dynamic monitoring, and (ii) colorimetric sensors, that enable direct visual detection of target analyte *via* obvious color changes. Among the latter, luminescent sensors are particularly advantageous for their ability to detect analytes intracellularly, serving as a powerful tool for studying Pd interference in biological systems. On the other hand, colorimetric sensors are highly appealing due to their simplicity, negligible capital-cost, and the suitability for on-site and real-time analysis.[40]

Since 2000, a large number of $Pd^{2+}$-responsive optical sensors have been developed encompassing small molecular organic/inorganic sensors,[41–44] MOFs (metal-organic frameworks),[45–49] and nanomaterials-based sensors.[50–53] Among these, metal-based luminescent probes have emerged as superior alternatives compared to purely organic counterparts for the selective detection of $Pd^{2+}$ ions. Unlike traditional organic probes, which often suffer from limitations such as small Stokes shifts, lower photostability, and significant spectral overlap, metal-organic systems offer a range of optical and physicochemical advantages that are highly desirable for sensing applications.[54–59] Such probes typically exhibit large Stokes shifts, enabling a more effective separation of excitation and emission wavelengths, thereby minimizing spectral interference and enhancing signal clarity.[60,61] Furthermore, they demonstrate high photostability and long luminescence lifetimes, allowing for prolonged monitoring with reduced background noise.[62–64] Beyond their favourable optical features, metal-based probes possess intrinsic structural versatility and tunability. Their modular architectures allow for the fine adjustment of electronic properties to achieve selective recognition of $Pd^{2+}$ ions, a task that is often more challenging with rigid organic fluorophores.[65] In addition, metal-organic systems frequently exhibit enhanced redox stability, magnetic behavior, catalytic activity, and increased solubility in aqueous environments- characteristics that further expand their utility in complex environmental matrices.[62,63,66,67] Crucially, the coordination chemistry of metal centers provides unique binding sites for $Pd^{2+}$ ions, enabling highly selective detection even in the presence of competing metal ions. This contrasts with many organic probes that rely predominantly on weak, non-specific interactions, resulting in compromised selectivity under real-world conditions. Collectively, these features have positioned metal-based luminescent probes at the forefront of sensor development, facilitating their integration into multifunctional platforms for environmental monitoring, biomedical diagnostics, and industrial process control.[68,69] Against this backdrop, the present review aims to systematically discuss recent advances in the design principles, sensing mechanisms, and application landscapes of luminescent and colorimetric metal-organic receptors specifically tailored for recognition of $Pd^{2+}$ ions in environmental matrices.

To date, $Pd^{2+}$ detection exclusively utilizing metal-based probes (such as metal complexes and MOFs) has received limited attention. Though some recent reviews have explored the optical detection of $Pd^{2+}$, they primarily focus on organic-based fluorophores or specific probe types like ratiometric or near infra-red (NIR) active systems.[20,23,32,46,49,52,70,71] The present study addresses this research gap by introducing key detection methods for $Pd^{2+}$ ions and summarizing the design and features of luminescent and colorimetric sensors. It compiles an updated systematic inventory of $Pd^{2+}$-responsive metal-based receptors reported over *ca.* 25 years, categorized *via* their structural complexity from metal complexes to MOFs.

## 2. Key characteristics and impacts of $Pd^{2+}$

Palladium is one of the most prominent elements employed across various industrial and chemical processes e.g., catalysis, automobiles, fuel cells, dental crowns, electric and electronic components.[72–74] Its utility is particularly significant in chemical, pharmaceutical and petroleum industries, where Pd-catalysts facilitate the syntheses of various drugs, natural products, polymers and other functional materials.[20,75,76] Such a widespread usage and related produced waste have increased $Pd^{2+}$ contamination into soil and water sources, and lead to severe effects on living beings. Indeed, early investigations clearly indicate that $Pd^{2+}$ can easily be transported to the biological materials, and thus, it accumulates in food chain and may cause potential health hazards.[18,77,78]

Over the years, there has been a remarkable growth in the research works focusing on Pd toxicity and related harmful effects on our ecological system. This reflects rising concerns about the increasing employment of Pd in various industrial and catalytic applications. An analysis of the trends from accessible studies on Pd-based catalysis and toxicity, as documented from the '*Web of Science*' database (Fig. 1), underscores an alarming pattern for Pd-based adverse impacts. Although the physiological roles and mechanistic pathways related to Pd-homeostasis have been explored in depth, the toxicological outcomes due to Pd exposure are gradually becoming evident. Excessive Pd intake (> 20.0 µg per person per day) has been linked to asthma, hair-loss, dermatitis, spontaneous abortion, and many other severe health issues in human beings.[70,79] These issues primarily arise from its ability to display strong binding with thiol-based amino acids, proteins (such as casein, silk fibroin, etc.), DNA and other bimolecular targets, leading to the damage of many cellular processes.[70,80,81] Hence, the recommended dietary intake of $Pd^{2+}$ is limited to < 1.5 µg to 15.0 µg per day per person, with an acceptable Pd threshold in the active pharmaceutical ingredients set as 5.0-10.0 ppm.[82]

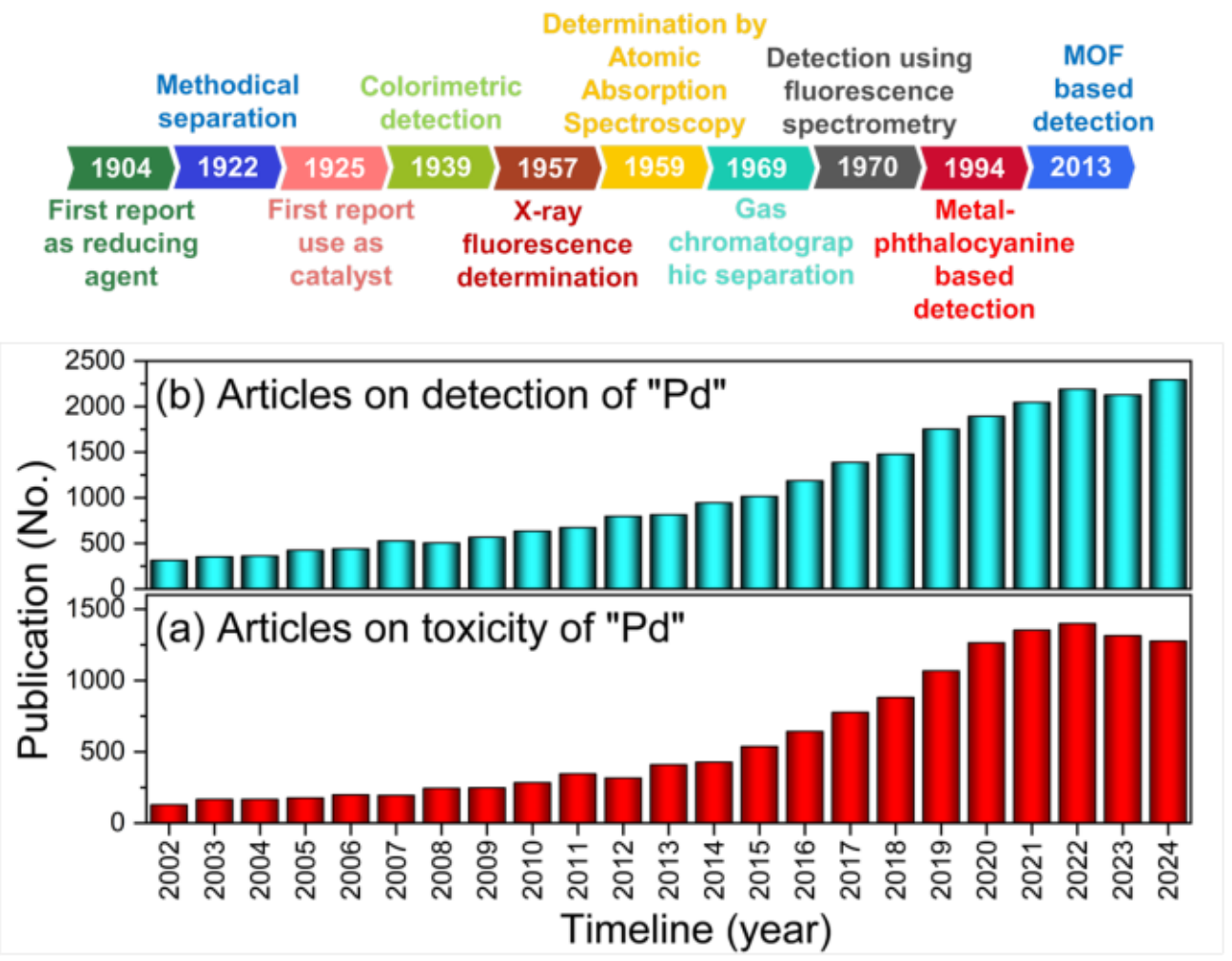

**Fig 1**. Comparative analysis of publications focused on (a) Pd toxicity and (b) Pd detection. Data adapted from ISI Web of Science, Feb 2025.

Furthermore, the toxic effects of Pd on plants cannot be overlooked. Its accumulation in water bodies due to industrial discharge is posing a threat to agricultural productivity and the food security (Fig. 2).[79,83] These findings highlight the urgent need and immediate action for mitigating the Pd-toxicity related risks. The employment of safer and more eco-friendly alternatives to Pd catalysts must be prioritized. Also, advancements of effective remediation and other clean-up approaches such as adsorption, phytoremediation, catalytic reduction, and the use of functionalized nanoparticles, to reduce the long-term Pd-induced impacts is essential to protect both humans as well as environment.[84–86] Contextually, optical sensing methods are emerging as remarkably promising tools to detect toxic heavy metals in complex environmental and biological matrices. These methods demonstrate huge potential for application in diverse systems, including live cells, tissues, and even animal models.[87–90] Their high sensitivity, selectivity, outstanding performance, reproducibility, and practicality position them as effective solutions for monitoring and managing Pd contamination in both environmental and biological settings.

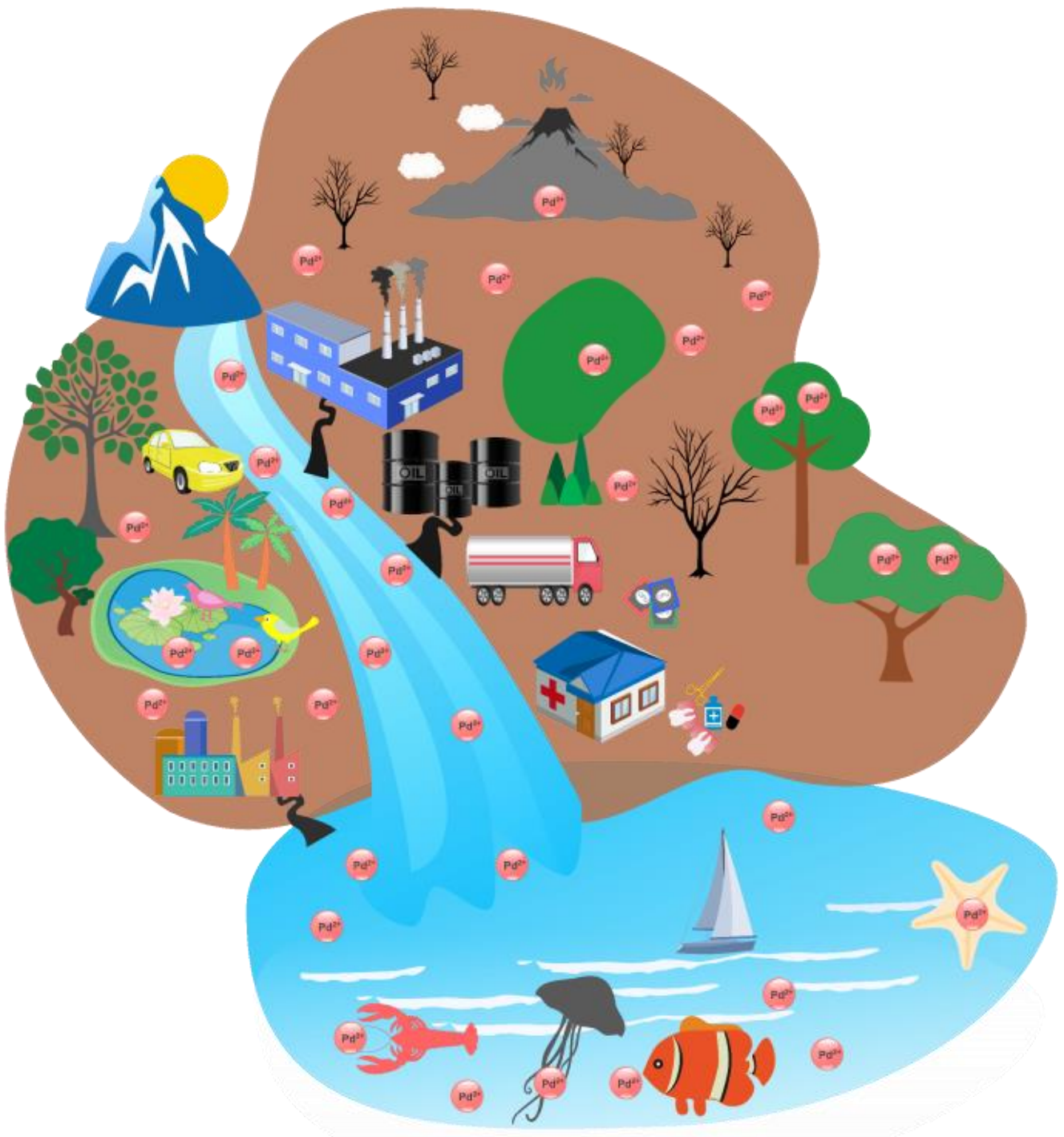

**Fig 2**. A schematic showing possible exposure pathways of $Pd^{2+}$ to living beings and environment [79,83].

## 3. Optical probes for $Pd^{2+}$ detection: Design principles

Metal ion responsive optical probes (luminescent or colorimetric) are typically designed by combining a chromophore- such as an organic dye or inorganic complex- with a receptor that specifically binds the target metal ion. This binding event triggers a measurable change in properties like luminescence or absorbance, serving as a signal for presence of the target metal.[35,91] The signalling efficacy of such probes directly relies on precise interaction, either intermolecular or intramolecular, between the receptor and the signalling unit.[92,93] Luminescent probes frequently operate through various photoinduced electron/energy transfer processes. Some key mechanisms utilized in their design include photoinduced electron transfer (PeT),[94–96] photoinduced charge transfer (PCT),[97,98] and fluorescence resonance energy transfer (FRET) events.[99–101] In addition, other mechanisms like metal-induced inhibition of excited-state-intramolecular-proton-transfer (ESIPT)[102,103] and aggregation-induced emission (AIE)[104–106] have also been efficiently utilized, broadening the scope of such luminescent probe's design. Each of these mechanisms offers distinct advantages, such as sensitivity or selectivity, and imposes certain restrictions related to environmental conditions or structural requirements. Through a systematic evaluation of various optical sensors, we observed differing optical responses of metal-based sensors to target analytes. It has been noticed that metal complexes and MOFs generally utilize the PeT mechanism for luminescence-based $Pd^{2+}$ detection. On the contrary, colorimetric probes typically rely on the intramolecular charge transfer (ICT) process, where the electronic charge shifts within the molecule in response to the target analyte, resulting in a color change rather than a luminescent signal.[40]

The design of a receptor (or binding unit) plays a critical role in heavy metal ions responsive probes, determining their ability to selectively detect a specific target ion over other competing species. To optimize the receptor-metal interaction and to enhance the selectivity, several factors must be carefully considered: *(i) the size and shape of the receptor's binding pocket*- the ionic radius of $Pd^{2+}$ ranges from 61.5 to 85.0 pm,[107] meaning that a $Pd^{2+}$-responsive probe can be designed to accommodate this range. This ensures that the designed probe is more likely to bind $Pd^{2+}$ while minimizing interference from other larger or smaller cations. However, the similarity in ionic radii between $Pd^{2+}$ and $Pt^{2+}$ makes selective detection challenging, as the receptor must be fine-tuned to differentiate between these two ions; *(ii) geometry of analyte-receptor adduct*- $Pd^{2+}$, with its $d^8$ electronic configuration, typically forms square planar complexes in which the highest energy $d_{x^2-y^2}$ orbital plays a dominant role to bind four ligands. The remaining *d* orbitals, which are of lower energy, are filled with the eight *d* electrons. Consequently, the four ligands align with the lobes of the $d_{x^2-y^2}$ orbital, resulting in the formation of stable square planar complexes. These complexes tend to exhibit low-spin characteristics, as the energy required for electron pairing in the *d* orbitals is lower than the energy associated with orbital splitting. This preference for low-spin configurations contributes to the stability and distinctive properties of $Pd^{2+}$ complexes, influencing the design of receptors with enhanced selectivity[108,109] and *(iii) nature of the donor atoms in receptor*- the Hard and Soft Acids and Bases (HSAB) principle can be implemented to design a probe ensuring the interaction between target ion and the receptor. $Pd^{2+}$, being a soft acid, shows high binding affinity with soft bases, such as thiols, thioethers (-SH and RSR'), S-bonded thiocyanates, sulfides, carbanions, phosphines ($PR_3$), terminal or bridged alkenes/alkynes, cyanides, as well as borderline bases like pyridine, bromide, nitrite, sulfite, and aromatic amines.[20,110] Therefore, to maximize the probe's selectivity for $Pd^{2+}$, the probe's receptor must include donor sites derived from the above-mentioned soft or borderline bases.

Though it is challenging to design a receptor that can easily distinguish between $Pd^{2+}$ and $Pt^{2+}$ due to their similar chemical properties, careful selection and integration of functional groups can induce detectable optical changes, such as variations in color, luminescence, or absorbance. By fine-tuning these factors, optical probes can achieve high sensitivity and selectivity for target metal ions. Achieving this requires attention to other factors, including solvent, pH, and the composition of the sensing medium, as competing or interfering species can significantly impact performance. Even minor variations in pH or solvent conditions may alter or enhance the probe's detection capabilities.[111]

*In fine*, this review highlights recent advancements, limitations, and the state-of-the-art in $Pd^{2+}$-responsive probes, encompassing luminescent and colorimetric metal-based receptors and their photophysical properties. Key studies are organized by complexity, progressing from metal complexes to MOFs, and provide insights into strategies for optimizing these sensory systems. While metal complexes exhibit superior solubility and stability, MOFs offer distinct advantages, including tunable pore sizes, high surface area, and structural flexibility, enabling simultaneous detection and adsorption of analytes. The transition from metal complexes to MOFs broadens detection capabilities, especially for analytes requiring tailored porosity or multifunctional designs. However, MOFs often face challenges such as limited stability, solubility in aqueous environments, and framework integrity upon ion incorporation, which constrain their practical applications.

## 4. Metal complexes based probes

To date, a diverse range of metal complexes (including those of transition and inner-transition metals) has been extensively explored as luminescent and/or colorimetric sensors for heavy metal ions.[69,112–116] These complexes are highly valued due to their large Stokes shifts, excellent redox and photophysical properties, good photostability, and absorption/emission in the visible region. These sensors consist of a ligand system coordinated to a single metal centre, which acts as the luminophore or chromophore. The metallic core is usually connected to a receptor unit *via* σ- or π-bonds, and the critical aspect is the mechanism by which the metal complex senses target cation. Upon binding with the target ion, the optical properties of the complex- such as luminescence or absorbance- undergo a measurable change. In luminescent sensing, this interaction can lead to either a "turn-on" or "turn-off" response, depending on the specific design of the system. If the receptor is connected to the metal core through a π-linker, the cation binding may also result in emission shift, providing an additional layer of detection specificity. On the other hand, colorimetric sensors detect cations by inducing a visible color change upon analyte binding, making them ideal for rapid and straightforward detection.

The selectivity of a probe towards any metal ion markedly depends on its coordination ability to that specific metal ion. The design of a $Pd^{2+}$-responsive luminescent or colorimetric probe is gaining increasing attention of researchers (Table 1), despite the complexity related to the similarity among the same group elements, especially platinum. To do this, the probe's receptor must be functionalized particularly with $Pd^{2+}$-specific donor sites. To date, mainly two types of the luminescent probes have been designed for $Pd^{2+}$ recognition: (i) a binding based probe- in which a reversible, non-covalent binding of $Pd^{2+}$ induces a luminescence change in the probe, resulting in the form of an optical response with luminophore and (ii) a chemical reaction based probe- in which the binding of $Pd^{2+}$ first induces a chemical transformation to the probe which ultimately leads to provide an optical signal, due to the formation of a luminescent product.[43] The selectivity of a particular probe is determined by the specificity of chemical reactions. $Pd^{2+}$ is usually considered as a luminescence quencher, even with the luminophores exhibiting remarkably high quantum yields. This quenching may be ascribed to the complex interactions related to $Pd^{2+}$ ion such as ICT, paramagnetic, as well as heavy atom effects.[20,23] Due to a $4d^8$ electronic configuration of $Pd^{2+}$, the occurrence of photoinduced energy/electron transfer between luminophore and $Pd^{2+}$ is quite possible. Therefore, several $Pd^{2+}$-responsive luminescent probes have been designed based on the luminescence quenching induced by $Pd^{2+}$ binding.

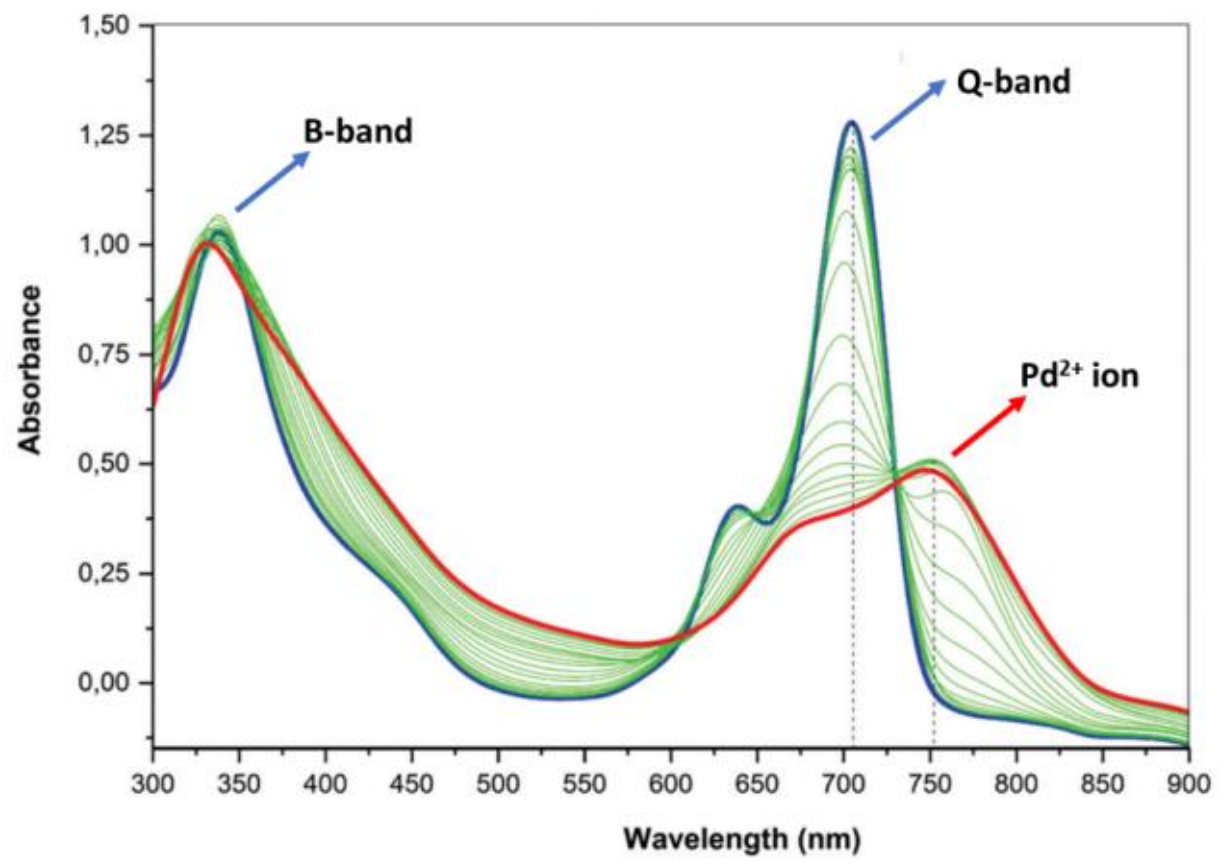

**Fig 3.** Typical UV-Vis spectral changes observed in metal-phthalocyanines upon introducing $Pd^{2+}$ ions. Adapted with permission from ref [42]. Copyright 2021, The Royal Society of Chemistry.

In year 1994, Gurol's and Gurek's groups reported the first generation transition metal complexes for $Pd^{2+}$ recognition.[117,118] These complexes contained a substituted phthalocyanine unit as the ligand scaffold. Since then, a diverse panel of metal-phthalocyanines (MPcs) notably bearing soft donor sites (e.g., S-containing groups) have been produced in this field.[42,119–127] Most of the studies on transition metal complexes serving as $Pd^{2+}$-responsive probes are based on the MPcs having $Mn^{2+}$, $Zn^{2+}$, $Co^{3+}$, $Cu^{2+}$ or $Ru^{2+}$ metal core.[119–127] More recently though, complexes involving lanthanides ($Ln^{3+}$) have also emerged as the exciting alternatives.[43] This review article majorly covers the important works related to this field with an emphasis laid down to the mechanistic routes of $Pd^{2+}$ sensing involved herein.

Phthalocyanines have attracted significant attention from the scientific community due to their extensive delocalized 18-$\pi$ electron system, enabling their versatile applications in catalysis, biomimetic chemistry, molecular electronics, photodynamic therapy (PDT), dye-sensitized solar cells (DSSC), molecular sensors, etc.[128–136] Over the past few years, Kandaz and co-workers [119–127,137] reported the use of various MPcs for luminescent or colorimetric detection of $Ag^+$ and $Pd^{2+}$ ions. The absorption spectra of phthalocyanines have been interpreted using Gouterman's four orbital models.[138] Typically, a phthalocyanine derivative shows characteristic Q-band of phthalocyanine ring at 600-750 nm due to $\pi \rightarrow \pi^*$ transition between its HOMO ($a_{1u}$ and $a_{2u}$) and LUMO ($e_g$). A high energy B-band at 300-400 nm is also observed which may be assigned to the deeper $\pi \rightarrow \pi^*$ transitions with a $D_{4h}$ symmetry.[139] In addition, one aggregation shoulder can be observed at slightly higher energy side of Q-band (blue-shifted). Notably, the presence of an electron-donating thioether moiety on the periphery of phthalocyanine often results into the red-shifted band as compared to its oxyether substituted moiety. Increase in the concentration of the phthalocyanine derivative causes aggregation through $\pi$–$\pi$ stacking interaction which ultimately leads to the decrease in the blue-shifted Q-band intensity.

It is well documented that a phthalocyanine moiety bearing an appropriate substituent (e.g., S-containing donor site) displays optical sensitivity toward precious as well as soft metal ions such as $Pd^{2+}$ and $Ag^+$.[117,118] The interaction of functionalized phthalocyanines with such soft metals significantly alters their optical properties, particularly the Q-band absorptions. These changes are readily observable through absorption and emission spectral analysis (Fig. 3). Phthalocyanines typically exhibit two types of aggregation effects (i) face-to-face *H*-aggregates and (ii) head-to-tail *J*-aggregates, depending on the nature and position of the functionality attached to the periphery. Unlike *H*-type aggregates, *J*-type aggregates are well known for their photoactive characteristics.[125,140] Both *H*- and *J*-aggregates directly correspond to hypsochromic and bathochromic shifts, respectively, in their UV-Vis spectra.

In an elegant report (in year 2007), Kandaz *et* al.[119] introduced a series of MPcs probes, designated **1-1d**, employing metallic cores $Ni^{2+}$, $Zn^{2+}$, $Co^{2+}$ and $Cu^{2+}$, respectively (Fig. 4(a)). Phthalocyanine rings were functionalized with reactive propane-1,2-diolsulfanyl groups to improve the solubility of resulting MPcs in polar solvents and facilitate the binding with target cation. UV-Vis spectra of **1a-1d** exhibited typical phthalocyanine based Q- and B- absorptions. The high-intensity Q-band absorptions for **1a-1d** have been observed at 688 nm, 682 nm, 672 nm and 679 nm, respectively. A weak absorption shoulder could also be observed at slightly higher energy side adjacent to the Q-band for each MPc. The derivatives featuring sulfanyl groups on the periphery displayed their Q-band absorption slightly shifted to the lower energy side, most likely due to the presence of electron-releasing thioether groups. On the other hand, the Soret $\pi \rightarrow \pi^*$ band exhibited only minor shift in all the cases. Increasing the phthalocyanine concentration resulted into the molecular aggregation which ultimately led to a blue-shift in Q-band with decreased absorption intensity. The *tetra*-substituted derivatives exhibited stronger aggregation features as compared to the *octa*-substituted ones. A combination of S and O donor sites present in the receptor contributed to a high affinity of probes towards heavy metal ions. The metal ion sensing ability of **1a-1d** has been investigated in DMF-THF solutions (1:9, *v/v*) (probe concentration ~$10^{-5}$ M). Absorption titration of probe **1c** with $Pd^{2+}$ induced notable spectral changes. The weak shoulder centred at 615 nm vanished completely, and a new band could be seen at relatively longer wavelength. Meanwhile, the Q-band absorption at 672 nm shifted slightly in presence of $Pd^{2+}$ ion. These spectroscopic observations suggested the perturbation of aggregation-disaggregation equilibrium by $Pd^{2+}$ ions.

To further improve the solubility of MPcs in polar solvents, Kandaz and co-workers (in 2007)[120] substituted the phthalocyanine periphery by thiohexylalcohol groups serving as soft bases (Fig. 4(c)). Probes **2a**, **2b** and **2c** containing $Zn^{2+}$, $Cu^{2+}$ and $Co^{2+}$, respectively, have been explored for cation sensing ability towards $Ag^+$, $Pd^{2+}$, $Hg^{2+}$ and $Cd^{2+}$ in MeOH-THF (1:9, *v/v*) solutions. As anticipated, the S-donor sites present on periphery were optically responsive for these soft metal ions, and thus, absorption spectroscopic studies have been carried out to monitor the sensing ability of probes **2a-2c**. UV-Vis spectra of **2a-2c** displayed the characteristic Q-band absorptions at 686 nm, 686 nm and 674 nm, respectively. The addition of $Pd^{2+}$ induced the disappearance of Q-bands while new bands at 654 nm (for **2a**), 655 nm (for **2b**) and 640 nm (for **2c**), attributed to oligomeric aggregated species, could be noticed. Though the intensity of B-bands remained almost unaltered, a blue-shift of 15-20 nm was observed for these bands across all three probes. Notably, the titration of **2a-2c** with $Pd^{2+}$ resulted into various isosbestic points which could be assigned to the formation of probe-$Pd^{2+}$ adducts along with the dynamic equilibrium between aggregation and disaggregation events. A high concentration of $Pd^{2+}$ reduced the intensity of Q-bands of both monomeric as well as oligomeric aggregated species with a moderate increase in the band associated with $n \rightarrow \pi^*$ transition (non-bonding S atoms to phthalocyanine $\pi^*$ orbital) at 400-500 nm. Furthermore, the solution's color immediately turned to dark green from blue-green. $Hg^{2+}$ and $Cd^{2+}$ failed to produce any absorption changes, but the addition of $Ag^+$ led to produce spectral profiles similar to those observed for $Pd^{2+}$ ions.

In the subsequent year (in 2008), the same group[121] prepared $Zn^{2+}$ and $Co^{2+}$ containing MPcs **3a** and **3b**, respectively, by replacing thiohexylalcohol group in probes **2a-2b** with (thiophene-2-carboxylate)-hexylthio group (Fig. 4(c)). The absorption spectra of **3a-3b** (~$10^{-4}$ M) exhibited the characteristic single high-intensity Q-bands at 688 nm (for **3a**) and 675 nm (for **3b**) when measured in THF-MeOH (5:1, *v/v*) solutions. Introduction of $Ag^+$ and $Pd^{2+}$ induced a sharp color change in the probe's solution from blue-green to green. Further addition of these ions resulted into a green solid, suggesting the formation of a new species. Moreover, the intensity of the Q-bands at 688 nm and 675 nm declined with concomitant increase in the bands at 645 nm and 627 nm (attributed to the oligomeric species). The B-band also slightly shifted to higher energy around 5-20 nm. The observed spectral behaviors were attributed to the formation of square planar **3**-$Pd^{2+}$ species, arising from the preferential binding of the soft S-donor sites with the soft $Pd^{2+}$ ion. This interaction led to the disruption of the aggregates afforded by the planar phthalocyanine molecules.

In year 2009, Kandaz's group[122] further constructed a series of non-peripherally substituted MPcs **4a**, **4b** and **4c** containing $Zn^{2+}$, $Cu^{2+}$ and $Co^{2+}$ metal cores, respectively (Fig. 4(b)). The phthalocyanine ring in these probes has been substituted non-peripherally by 6-hydroxyhexylsulfanyl group to improve the probe's solubility in polar solvents as well as to enhance the metal ion sensing abilities. UV-Vis spectra of **4a-4c** displayed typical Q-band absorptions centred around 695-710 nm arising from the phthalocyanine-based $\pi \rightarrow \pi^*$ transition. In addition, the deeper $\pi \rightarrow \pi^*$ transitions gave rise to the B-band absorptions near 300-400 nm. Notably, the Q-bands in the non-peripherally substituted **4a-4c** were found to be red-shifted (upto 20 nm) compared to the corresponding peripherally substituted MPcs (i.e., **4a-4c**) in the same solvent systems. UV-Vis spectroscopic study was further employed to investigate the cation sensing ability of **4a-4c** in THF-MeOH (9:1, *v/v*) solutions. The titration of **4a** and **4b** with soft metal ions, such as $Ag^+$ and $Pd^{2+}$, led to an obvious color change from blue-green to dark-green (visible to the naked eye). Upon gradual addition of $Pd^{2+}$ to **4a** and **4b**, the corresponding high-intensity bands at 710 nm and 708 nm declined in the intensity. Simultaneously, new bands emerged at 662 nm (in **4a**) and 668 nm (in **4b**), attributed to the increasing concentration of oligomeric aggregated species in the solution.

With a focus again on the peripherally functionalized derivatives, a series of MPcs **5a-5e** employing $Pb^{2+}$, $Zn^{2+}$, $Cu^{2+}$, $Co^{2+}$ and $Mn^{3+}$, respectively, was presented by the same group[123] in year 2010 (Fig. 4(a)). In this study, the phthalocyanine ring was substituted by 6-hydroxyhexan-3-ylthio functionality to enhance the selective binding for soft metal ions. A single high intensity Q-band absorption could be observed at 726 nm, 700 nm, 693 nm, 679 nm and 755 nm for **5a**, **5b**, **5c**, **5d** and **5e**, respectively, in THF solutions. The spectral data clearly indicated that the Q-band absorptions were highly sensitive for various substituents present on the periphery of phthalocyanines. Functionalization of S-containing groups over macrocyclic benzenoid remarkably influenced the optical characteristics of resultant phthalocyanines. As a consequence, UV-Vis spectra of **5a-5e** significantly red-shifted, underscoring the potential of these MPcs for bio-medical applications. The cation-sensing ability of **5a-5e** was monitored using UV-Vis analyses in THF-MeOH solution (9:1, *v/v*). Titration with $Pd^{2+}$ ions induced substantial changes in both Q- and B-bands of probes **5a-5e**, which could be attributed to the formation of thioether-$Pd^{2+}$ adducts. These interactions led to the formation of dimeric species, ultimately resulting into planar aggregates.

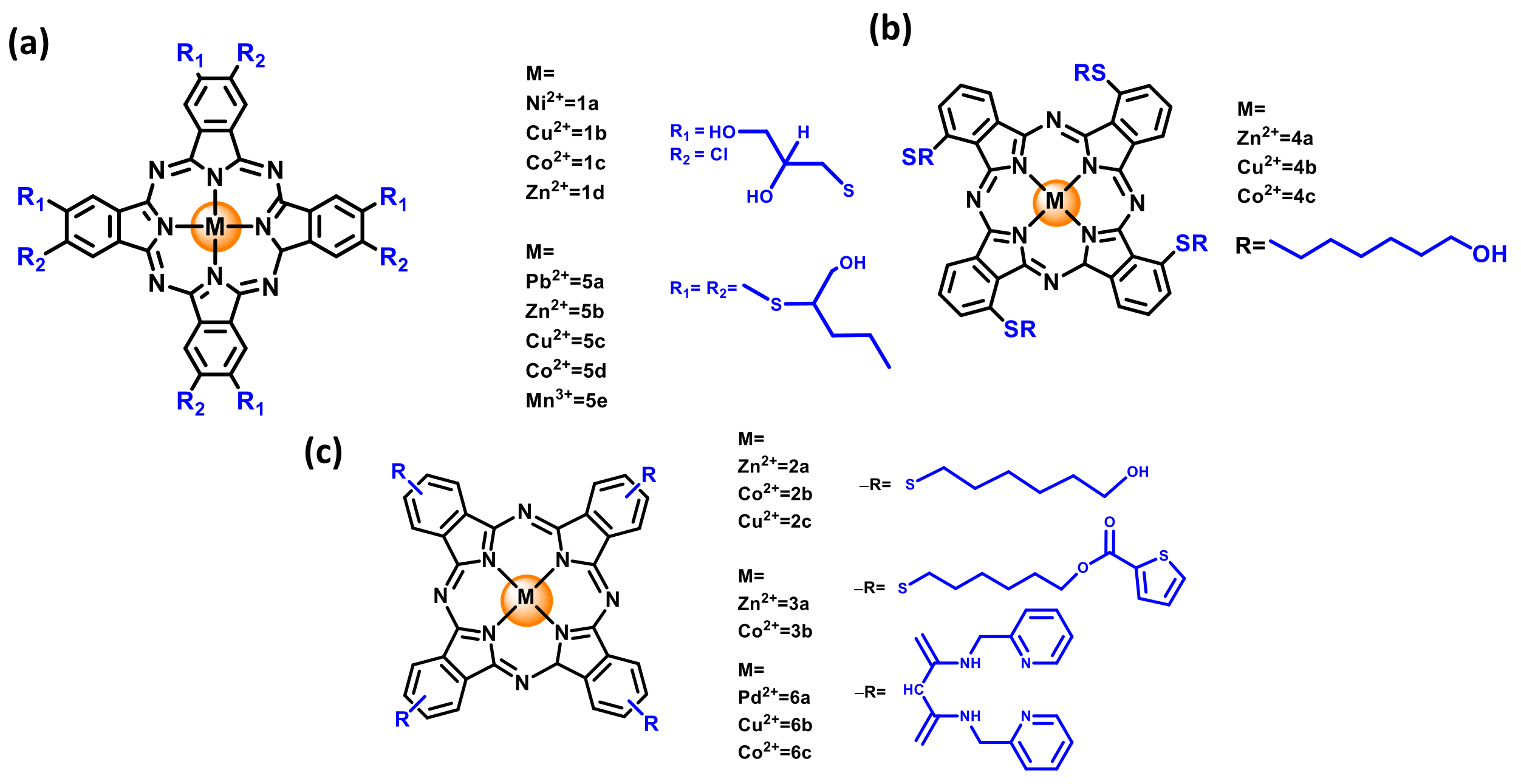

**Fig 4.** Chemical drawings of metal-phthalocyanines probes **1-6**.

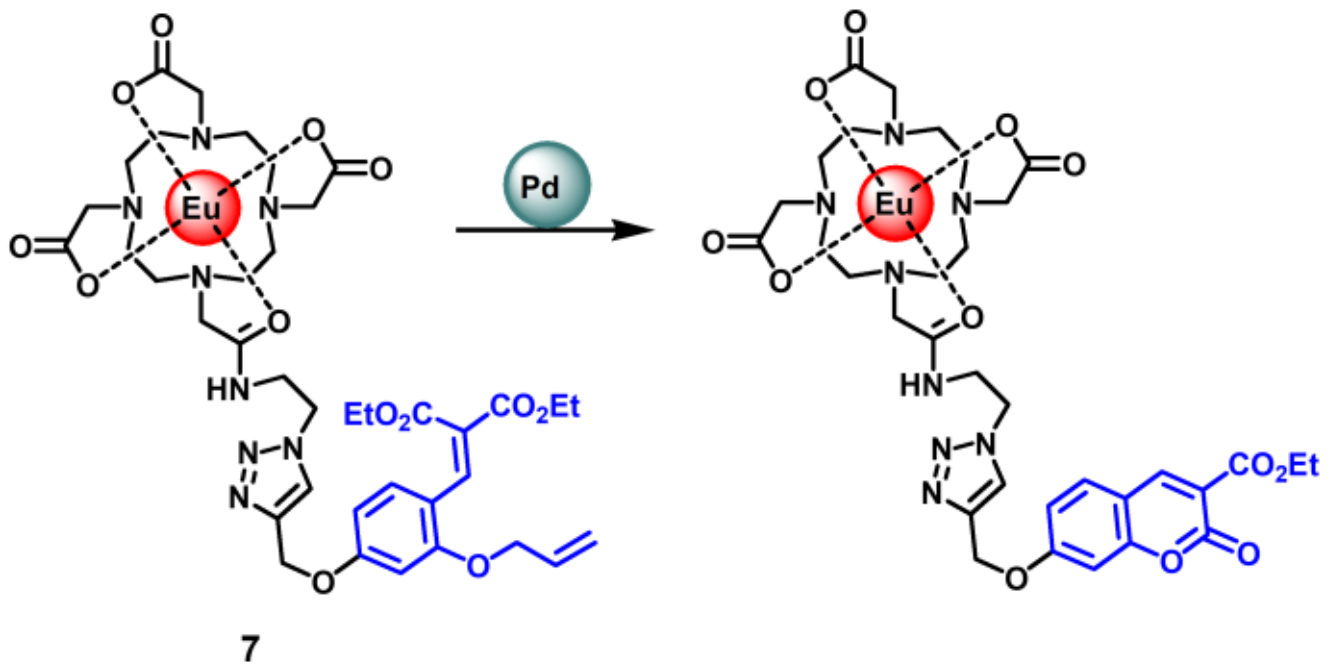

**Fig 5.** Reaction-based sensing of Pd(0)/(II) in presence of $PPh_3$ by $Eu^{3+}$-based probe **7**.

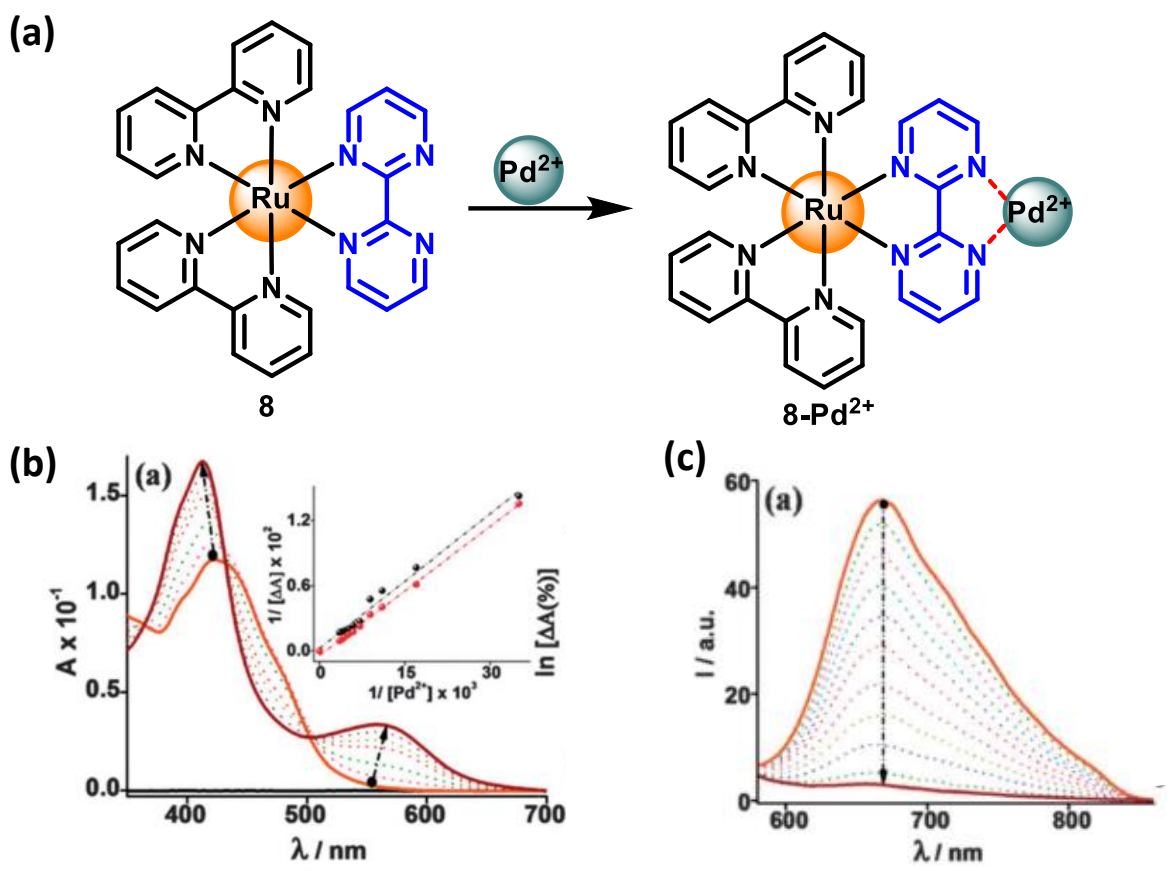

**Fig 6.** (a) Binding mode of $Ru^{2+}$-based probe **8** with $Pd^{2+}$ ion; (b) UV-Vis spectra of **8** after the addition of $Pd^{2+}$ and (c) Luminescent quenching in **8** upon $Pd^{2+}$ addition in water ($\lambda_{ex}$= 428 nm). Adapted with permission from ref [145]. Copyright 2014, The Royal Society of Chemistry.

In 2011, Kandaz's group[124] demonstrated a series of *tetra*-substituted amido based MPcs **6a-6c** (Fig. 4(c)) (M = $Pd^{2+}$, $Cu^{2+}$ and $Co^{2+}$). UV-Vis studies for **6a-6c** revealed the typical Q- and B-band absorptions near 600-700 nm and 300-350 nm, respectively. In addition, an aggregation shoulder was noticed at 624 nm, 632 nm and 638 nm for **6a-6c**, respectively. The inclusion of aza and amide groups as peripheral substituents notably caused a blue-shift in the Q-band. The spectrometric titration of probes **6a** and **6b** with $Pd^{2+}$ led to a decrease in the Q-band intensity at 663 nm and 674 nm, respectively, in THF-MeOH (5:1, *v/v*) solutions. Owing to a limited solubility of the resulting oligomers, no clear isosbestic points could be observed, and further addition of $Pd^{2+}$ led to the intensity reduction of all the absorption bands.

Lanthanide ($Ln^{3+}$) complexes are also emerging as appealing alternatives to the transition metal complexes in the field of $Pd^{2+}$ sensing, owing to their exceptional photophysical properties. Such complexes typically employ a suitable organic chromophore (i.e., antenna) which can transfer its energy to sensitize $Ln^{3+}$ ion. The triplet state of the antenna is designed to closely match the excited state of the $Ln^{3+}$, minimizing the energy loss due to non-radiative pathway and preventing energy transfer roll-back from the $Ln^{3+}$ back to antenna. This eventually results in an allowed *f-f* transition, a process known as 'antenna effect' or the ligand-sensitized emission.[38,141,142] Taking advantage of this process, in year 2012, Pershagen *et al.*[43] synthesized a $Ln^{3+}$-based probe **7** featuring an allyl cage for selective recognition of $Pd^0$ ions (Fig. 5). The originally non-emissive probe displayed a 'turn-on' luminescent response in presence of $Pd^0$ when excited at 356 nm. No other metal ions including $Zn^{2+}$, $Ru^{3+}$, $Rh^+$, $Pt^{2+}$, $Ni^{2+}$, $Mn^{2+}$, $Mg^{2+}$, $Li^+$, $K^+$, $Fe^{3+}$, $Fe^{2+}$, $Cu^{2+}$, $Co^{2+}$, $Au^+$, $Au^{3+}$ and $Ag^+$ enabled such behavior except $Au^{3+}$ in $CH_3CN$: pH 10.0 borate buffer (1:1). The $\pi$-phillic $Pt^{2+}$ ion also interacted with probe **7** while its response was much lesser as compared to $Pd^0$ ions. Interestingly, probe could also detect $Pd^{2+}$ in the presence of $PPh_3$.

Ruthenium(II)-polypyridyl complexes have also been extensively employed as luminescent sensors for sensing of metal ions because of their large stokes shift, excellent photophysical properties, photostability, and emission lying in the visible region.[143–148] Their physicochemical features can be fine-tuned by modifying the periphery around Ru(II) centre, either by altering the number of polypyridyl ligands or by changing their nature. Moreover, the octahedral geometry of such complexes can give access to 3D structures, improving their versatility in sensor's design.[149] In year 2014, Kumar *et al.*[145] synthesized a Ru(II) probe **8** for selective recognition of $Pd^{2+}$ in aqueous media (Fig. 6(a)). UV-Vis spectrum of **8** displayed an intense ligand-centred (LC) $\pi \rightarrow \pi^*$ transition at 283 nm coupled with a broad MLCT absorption at 428 nm. Addition of $Pd^{2+}$ (50 ppm) to **8** induced a hypochromic shift (*ca.* 1.8-fold) in the LC band whereas the MLCT band exhibited hypsochromic (upto 16 nm) and hyperchromic shifts (*ca.* 1.5-fold). Interestingly, a new peak appeared at 565 nm with three clear isosbestic points centred at 364 nm, 432 nm and 503 nm, suggesting the formation of a new species upon $Pd^{2+}$ inclusion (Fig. 6(b)). This 'turn on' and 'turn left' response was also followed by a colorimetric change from orange to dark red. Probe **8** was found luminescent in nature and its emission could be observed at 670 nm ($\lambda_{ex}$ = 428 nm). The addition of $Pd^{2+}$ to **8** induced the luminescence quenching response, ascribed to the paramagnetic nature of $Pd^{2+}$ ion (Fig. 6(c)). These data clearly indicated the formation of a new heterobimetallic Ru(II)-Pd(II) complex i.e., **8**-$Pd^{2+}$. This probe selectively detected $Pd^{2+}$ among other metal ions such as $Zn^{2+}$, $Pb^{2+}$, $Ni^{2+}$, $Hg^{2+}$, $Mn^{2+}$, $Ag^{+}$, $Au^{3+}$, $Cu^{2+}$, $Co^{2+}$, $Cd^{2+}$ and $Fe^{2+}$ as well as highly competing species from platinum group e.g., $Pt^{2+}$, $Rh^{3+}$ and $Ru^{3+}$. Job's plot analyses clearly indicated a 1:1 stoichiometric ratio for the formation of adduct **8**-$Pd^{2+}$. Moreover, **8** showed a very low detection limit for $Pd^{2+}$ (i.e., $1.54 \times 10^{-6}$ M), with a binding constant value of $1.89 \times 10^{3}$ $M^{-1}$. This probe was also tested for imaging $Pd^{2+}$ in real water specimens such as pool and tap water, as a proof-of-concept experiment. Other unknown $Pd^{2+}$ samples were also detected by **8** with an accuracy of over 90%.

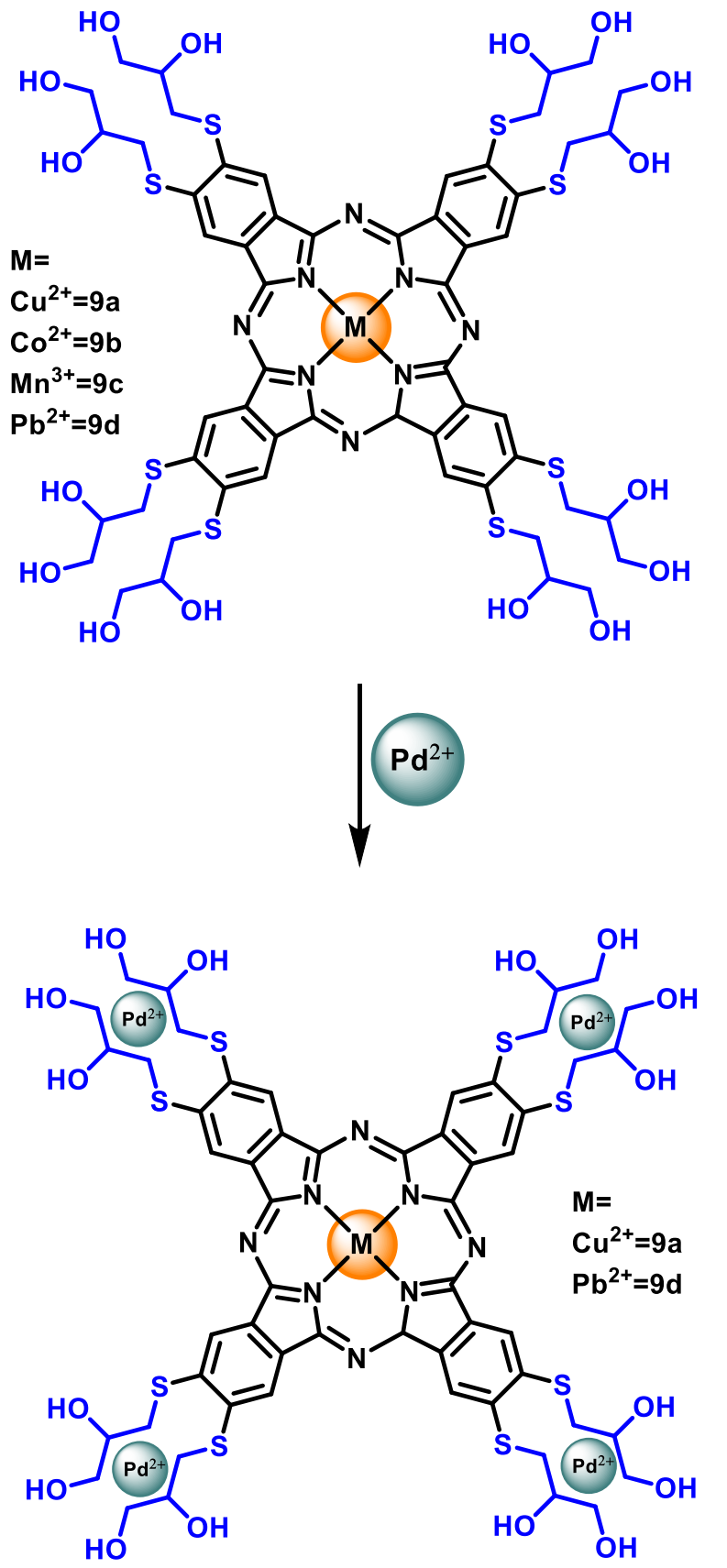

**Fig 7.** Chemical drawings of MPcs **9a-9d** and binding mode of $Pd^{2+}$ with **9a** and **9d**.

In an elegant study (in year 2014), Kandaz and group[125] developed a series of 'scorpion-like' MPcs **9a-9d** comprised of $Cu^{2+}$, $Co^{2+}$, $Mn^{3+}$ and $Pb^{2+}$, respectively (Fig. 7). The phthalocyanine periphery was substituted by 1,2-dihydroxyl-propylthio group serving as the receptor for target metal ions. The absorptions assigned to the Q-bands were found at 693 nm, 706 nm, 740 nm and 720 nm, respectively, for **9a-9d**. Additionally, a weak shoulder in the range 400-525 nm appeared in all the cases due to the characteristic charge transfer transition. UV-Vis study was employed to investigate the sensing ability of these probes for various metal ions like $Ag^{+}$, $Pd^{2+}$, $Zn^{2+}$, $Ni^{2+}$, $Na^{+}$, and $Mg^{2+}$ in THF-DMF (3:1, *v/v*) solutions. Amongst them, only $Ag^{+}$ and $Pd^{2+}$ enabled **9a-9d** to produce a noticeable change in the absorption spectra. With only a slight variation, the spectral changes were almost identical for $Ag^{+}$ and $Pd^{2+}$ ions. After aggregation of **9d** promoted by $Ag^{+}$, the absorption maximum at 720 nm declined with a concomitant increase in 678 nm band. Relatedly, the observed spectral changes have been assigned to the formation of aggregated dimer species between phthalocyanines (face-to-face *H*-aggregation). Addition of **9d** to $Pd^{2+}$ exhibited almost similar spectral profiles except that the aggregation bands were found at slightly lower wavelength in this case. The intensity of the B-bands also declined at 320-360 nm, especially for probes **9a** and **9d**. The *J*-type aggregation *via* head-to-head or head-to-tail approaches in **9b** and **9c** promoted substantial changes in the longer wavelength. The aggregation process led to a red-shift in the Q-absorption which was more pronounced in **9c** (*ca.*

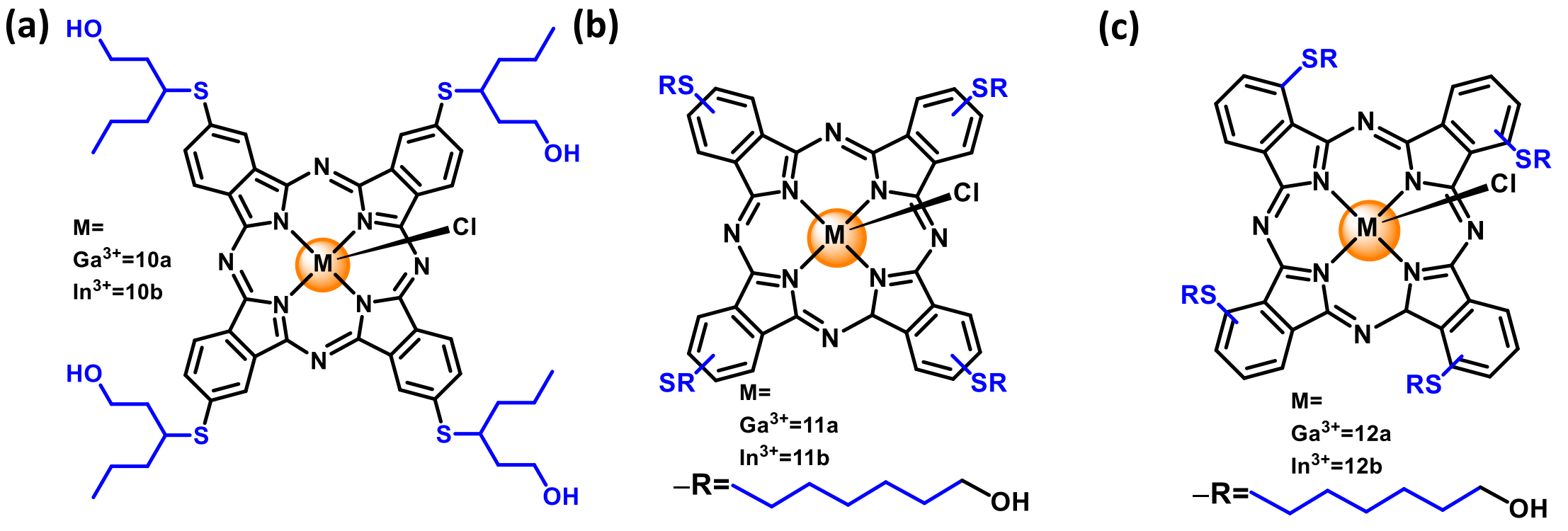

**Fig 8.** Chemical drawings of MPcs **10-12**.

15 nm) due to a relatively stronger guest-host interaction. The spectral titration was followed by gradual color change of probe's solution except in the case of **9b** and **9c**.

In 2015, Kandaz's group[126] utilized $Ga^{3+}$ and $In^{3+}$ incorporated MPcs **10a** and **10b**, peripherally *tetra*-substituted by 1-hydroxyhexan-3-ylthio groups (Fig. 8(a)). The trivalent metallic cores were deliberately chosen due to their strong axial substitution capability and pronounced diamagnetic features. Authors anticipated that the substituents present on the axial positions might induce a dipole moment perpendicular to the ring, influencing the spatial relationship among the molecules through some steric factors. The high-intensity Q-bands appeared at 707 nm and 708 nm for **10a** and **10b**, respectively. In presence of $Pd^{2+}$ and $Ag^{+}$, a color change from green to blue could be noticed in DMF solutions of **10a** and **10b** (*ca.* 10 mM). During the titration process, the absorption bands between 320 nm and 360 nm (B-bands) declined with a slight shift to the longer wavelengths. These spectral changes were ascribed to the increased concentration of dimeric *H*-type aggregates. Similarly, the addition of $Ag^{+}$ led to a decrease in the intensity of the both Q- and B- absorptions.

In the course of same year, Kandaz, Yarasir and group[127] reported another series of $Ga^{3+}$ and $In^{3+}$ (**11** and **12**) based MPcs containing peripheral and/or non-peripheral substituted phthalocyanines (Figs. 8(b) and (c)). UV-Vis spectra of MPcs displayed single high-intensity Q-bands centred at 704 nm, 710 nm, 734 nm and 740 nm, and the B-bands near 331 nm, 341 nm, 346 nm and 345 nm for **11a**, **11b**, **12a,** and **12b**, respectively. The emission maxima of MPcs in DMF solutions appeared at 723 nm, 734 nm, 754 nm and 728 nm for **11a**, **11b**, **12a** and **12b**, respectively, with the corresponding emission quantum yields ($\Phi_{em}$) as 0.2666, 0.0551, 0.2295 and 0.0692. The addition of $Pd^{2+}$ (~$10^{-3}$ M) to MeOH-THF solution of MPcs (*ca.* $10^{-5}$ M) (1: 9, *v/v*) could be monitored by recording their absorption spectral changes. The bands attributed to dimeric *H*-type aggregates along with the B-band (at 320-360 nm) exhibited an intense blue-shift. The binding ratio was calculated to be 3:1 for $Pd^{2+}$ with nearly similar changes observed for $Ag^{+}$ ions. The intensity of the Q- and B- bands diminished while the bands due to dimeric *H*-type aggregates showed enhancement.

In the succeeding year (in 2016), Kandaz and coworkers[137] synthesized a $Zn^{2+}$-phthalocyanine probe **13** having (3-hydroxynaphthalen-2-yl)-methylene amino group at the periphery for a selective detection of $Pd^{2+}$ ions (Fig. 9(a)). The characteristic Q-band in UV-Vis spectrum of **13** was observed at 705 nm while its emission maximum centred near 710 nm in THF solution. Upon interaction with $Pd^{2+}$, the Q-band of **13** significantly declined together with a red-shifted of *ca.* 9 nm. This spectral change has been attributed to the interaction of $Pd^{2+}$ with peripheral S-containing receptor, leading to a decrease in its HOMO-LUMO band gap. On the other hand, the peak at 372 nm blue-shifted to 328 nm with an increment in the absorption intensity. A new band also appeared at 435 nm, associated with the charge transfer transition. Upon excitation at 430 nm, the emission maxima of **13** have been measured at 525 nm and 723 nm. A significant quenching in the emission bands of **13** could be noticed upon addition of $Pd^{2+}$, typically due to the existence of *J*-type aggregation (Fig. 9(b)). Other cations such as $Zn^{2+}$, $Hg^{2+}$, $Pb^{2+}$ and $Ag^{+}$ notably failed to show any remarkable changes in the absorption profile of **13**.

Over past few years, rhodamine-based luminescent probes have also seen gradual growth in detection of $Pd^{2+}$ ions.[150–152] In year 2018, Huang *et* al.[153] developed a ferrocene-rhodamine derived probe **14**, and tested its sensing ability towards various cations such as $Li^{+}$, $Na^{+}$, $K^{+}$, $Cs^{+}$, $Mg^{2+}$, $Ca^{2+}$, $Sn^{2+}$, $Pb^{2+}$, $Pd^{2+}$, $Zn^{2+}$, $Co^{2+}$, $Ni^{2+}$, $Pt^{2+}$, $Tb^{3+}$, $La^{3+}$, $Ce^{3+}$, $Eu^{3+}$, and $Er^{3+}$ as well as anions such as $F^{-}$, $Cl^{-}$, $Br^{-}$, $H_2PO_4^{-}$, $HSO_4^{-}$, $AcO^{-}$, etc. in $H_2O$-THF (9:1, *v/v*) solutions (Fig. 10(a)). Probe **14** exhibited two characteristic absorption bands at 316 nm and 467 nm which could be attributed to the presence of ferrocenyl moiety. Absence of any absorption above 467 nm in the visible region clearly indicated the existence of **14** in the spirolactam form. Addition of $Pd^{2+}$ to **14** in $H_2O$-THF (9:1, *v/v*, 10.0 μM) solution produced a color change from yellow to pink accompanying the appearance of new band at 562 nm (Fig. 10(b)). This change has been ascribed to the opening of spirolactum ring to afford a fully delocalized xanthene type structure. The gradual addition of $Pd^{2+}$ resulted into the decrease in 316 nm band with a simultaneously enhanced intensity at 562 nm. The detection limit and binding constant values were computed as $8.46 \times 10^{-9}$ M and $4.0 \times 10^{6}$ $M^{-1}$, respectively, and a 1:1 binding stoichiometry was evidenced by Job's plot analysis. The free probe showed no luminescence in the range 565-690 nm when excited at 558 nm ($\Phi_{em}$ 0.004), suggesting the existence of spirolactum ring in the probe's structure. However, a 'turn-on' luminescence response upto *ca.* 82-fold ($\Phi_{em}$ 0.032) could be noticed at 582 nm after $Pd^{2+}$ addition, upholding the ring opening of the spirolactum moiety (Fig. 10(c)). Furthermore, the $Pd^{2+}$ selectivity of **14** was unaffected by other interfering cations and anions over a wide range of pH from 5.0 to 12.0. A coordination-based mechanism supported by FT-IR and DFT studies has been proposed in this study. For practical applications in real samples, **14** was exploited to monitor $Pd^{2+}$ in living cells by luminescence imaging studies. Live HeLa cells were incubated with probe **14** (50.0 µM) at 37 °C for 30 min. HeLa cells without $Pd^{2+}$ exhibited no luminescence; however, a strong luminescence could be realized upon addition of 100 µM $Pd^{2+}$ within 10 min, implying that **14** can be utilized for a quick intracellular $Pd^{2+}$ imaging in live cells.

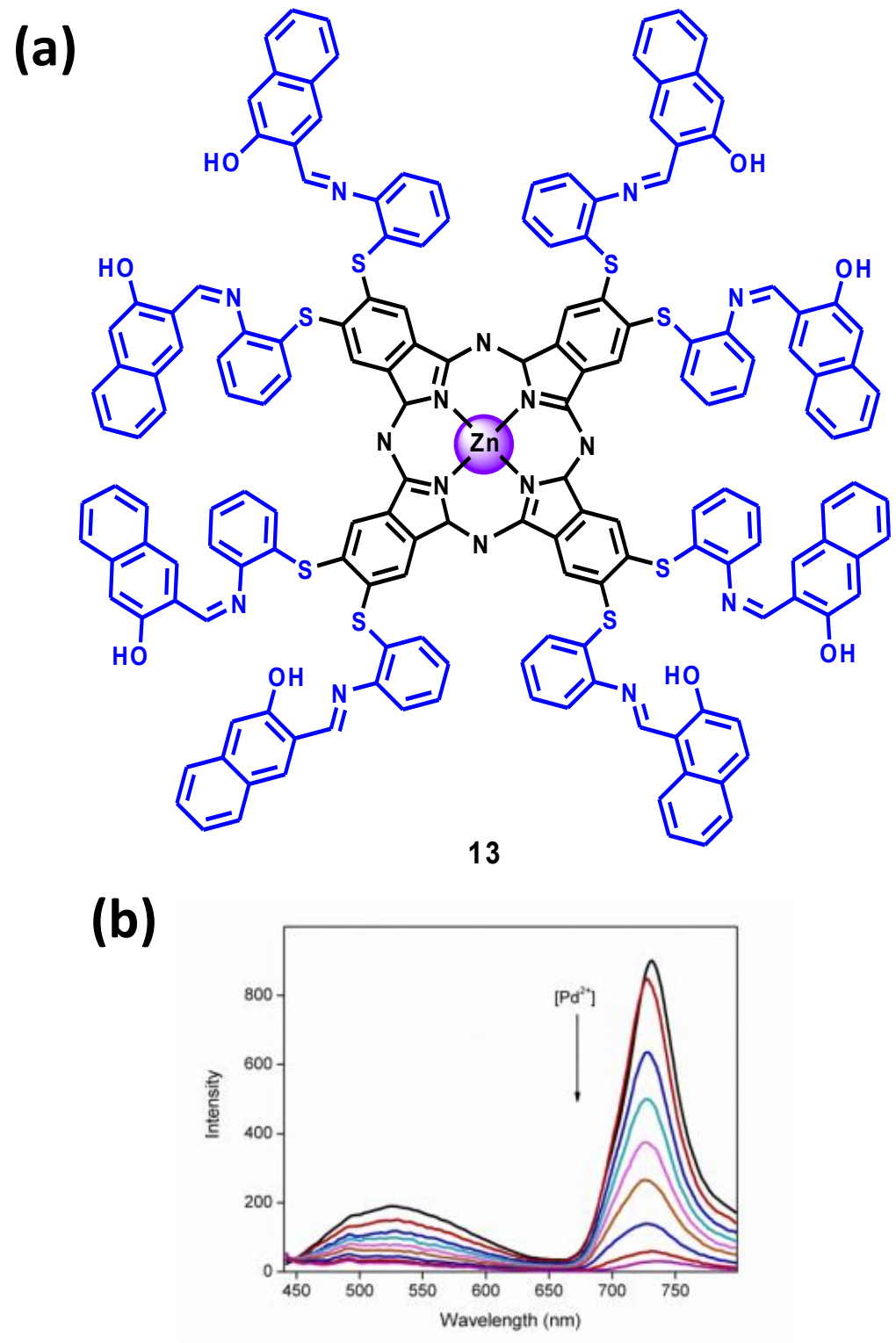

**Fig 9.** (a) Chemical drawing of $Zn^{2+}$-phthalocyanine probe **13** and (b) luminescent quenching in the emission band of **13** upon addition of $Pd^{2+}$ ion in THF solution ($\lambda_{ex}$= 350 nm). Adapted with permission from ref [137]. Copyright 2016, Elsevier.

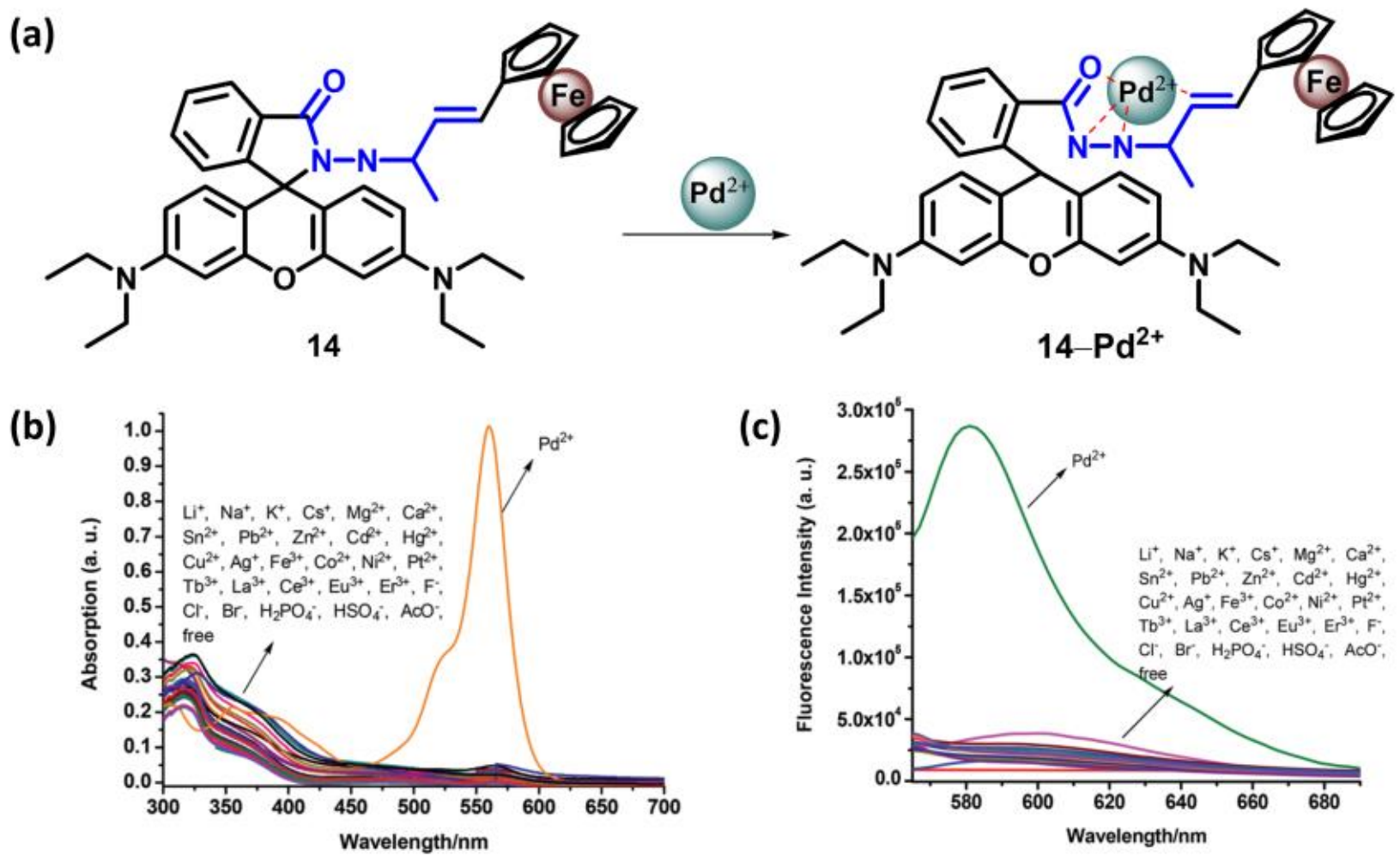

**Fig 10.** (a) Binding mode of $Pd^{2+}$ in ferrocene-rhodamine based probe **14**; (b) UV-Vis spectra of **14** in presence of different ions and (c) luminescence enhancement in **14** after the addition of $Pd^{2+}$ in THF-water (1:9, *v/v*) solution ($\lambda_{ex}$= 558 nm, pH = 7.3). Adapted with permission from ref [153] Copyright 2018, The Royal Society of Chemistry.

peripherally substituted by 4-bromobenzylthio group (Fig. 11). These probes could selectively detect $Pd^{2+}$ ions *via* the aggregation process in MeOH-THF (1:9, *v/v*) solutions, and the detection process was associated with a color change of probe's solution from green to light green. In UV-Vis spectra of **15**, the characteristic high-intensity Q-bands due to monomeric species could be observed at 686 nm (for **15a**), 687 nm (for **15b**) and 704 nm (for **15c**). Interaction of probes with $Pd^{2+}$ promoted the aggregation process, as evidenced by the decrement in original Q-bands with a concomitant increase at 642 nm, 646 nm and 670 nm, respectively, for the probes **15a**, **15b** and **15c**. The stoichiometric binding ratio for **15a** and **15b** was 1:2, while a 1:1.5 stoichiometry was depicted in **15c**. Notably, probe **15a** failed to exhibit any luminescent feature likely due to the presence of paramagnetic $Cu^{2+}$ ion. The emission maxima for

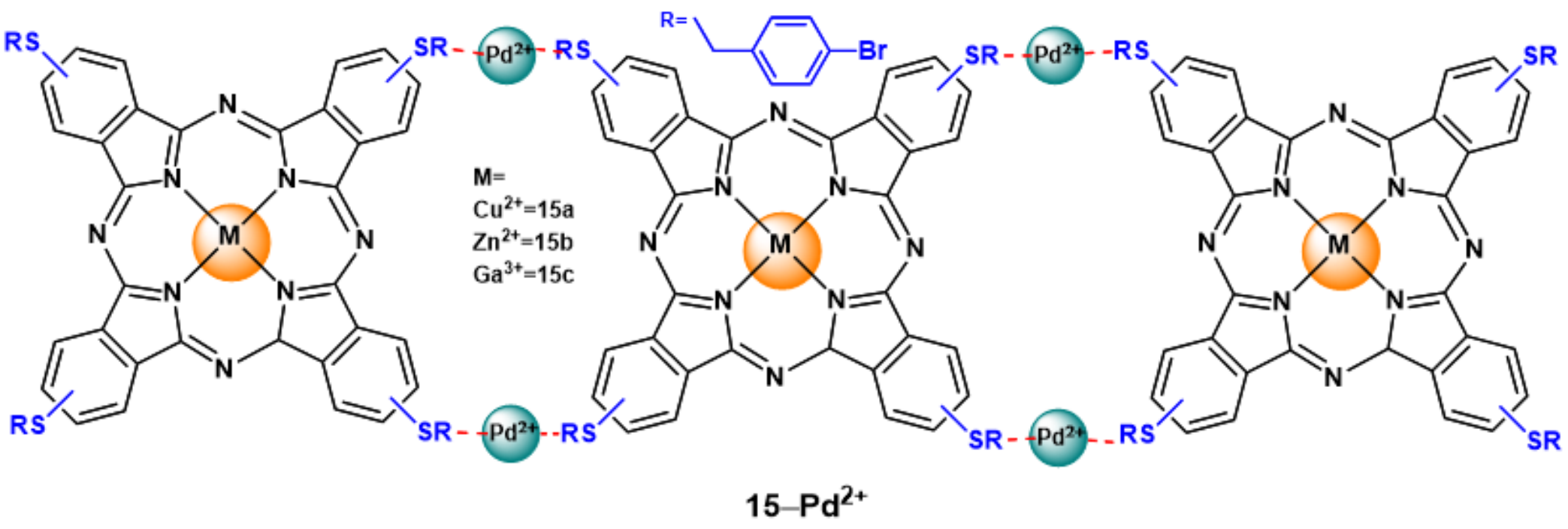

**Fig 11.** *H*-type aggregates formation in metal-phthalocyanines probe **15** upon $Pd^{2+}$ addition.

In year 2019, Bilgiçli's group[154] synthesized a series of MPcs probes **15a** (M = $Cu^{2+}$), **15b** (M = $Zn^{2+}$) and **15c** (M = $Ga^{3+}$), **15b** and **15c** were measured at 704 nm and 721 nm, respectively, with the corresponding emission quantum yields respectively, with the corresponding B-bands at 331 nm, 337 nm and 331 nm. Upon $Pd^{2+}$ addition, the intensity of Q-bands reduced significantly with the emergence of new bands at 756 nm for **16a**, 752 nm for **16b** and 758 nm for **16c**, suggesting the existence of *J*-type aggregation. The binding ratio between $Pd^{2+}$ and the probe was calculated to be 1:2. The emission maximum of **16b** lied near 722 nm that red-shifted to 731 nm in the presence of $Pd^{2+}$ ion. Based on the prior observations, this red-shift could be ascribed to a decrease in the HOMO-LUMO gap as a result of *J*-type aggregation. The Stern-Volmer coefficient for **16b** was calculated at $1.2 \times 10^5$ mol/L while the binding constant was depicted as $3.4 \times 10^7$ $M^{-1}$.

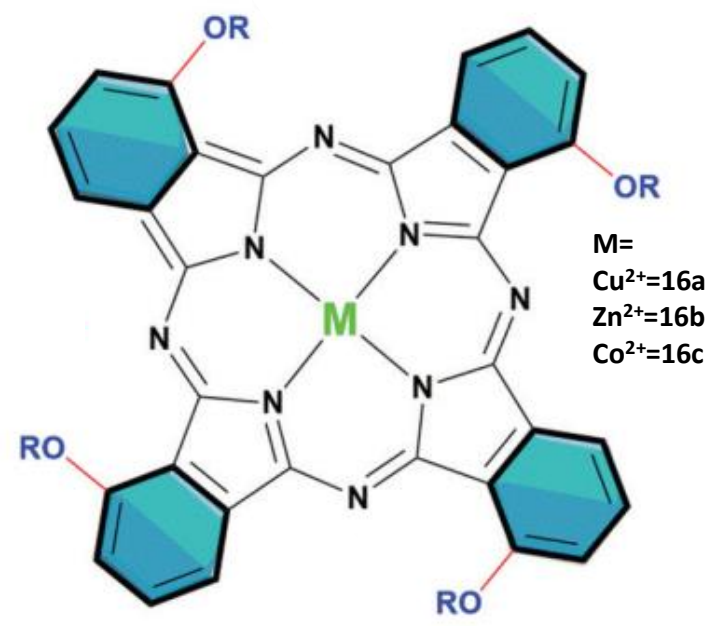

**Fig 12.** Chemical drawing of MPcs **16a-16c**. Adapted with permission from ref [42]. Copyright 2021, The Royal Society of Chemistry.

($\Phi_{em}$) of 0.14 and 0.16. The emission bands of **15b** and **15c** markedly quenched upon $Pd^{2+}$ addition due to the formation of *H*-type aggregates. These aggregates were non-luminescent in nature because of their lower state stabilization, resulting in the fast transition between the excited- and the ground state. As the dipole moment vanishes, most of the losses are non-radiative in nature. Other metal ions such as $Ag^+$, $Cd^{2+}$, $Cu^{2+}$, $Fe^{2+}$, $Hg^{2+}$, $Pb^{2+}$, $Ni^{2+}$ and $Zn^{2+}$ produced nominal changes in the spectral profiles of the probes.

In 2021, the same group[42] constructed MPcs **16a** (M = $Cu^{2+}$), **16b** (M = $Zn^{2+}$) and **16c** (M = $Co^{2+}$) that were non-peripherally substituted by 2,3-bis(hexadecylthio)propan-1-ol groups (Fig. 12). UV-Vis spectra of **16a-16c** in THF solutions displayed the characteristic Q-absorptions at 706 nm, 704 nm and 701 nm,

In 2021, Schwarze[41] reported two $Mg^{2+}$-based MPcs **17a** and **17b** bearing alkyl-thio groups in the phthalocyanine's periphery for selective detection of $Pd^{2+}$ ions (Fig. 13(a)). UV-Vis spectral analysis revealed a broad absorption band spanning 320-450 nm, overlapping with the characteristic $\pi \rightarrow \pi^*$ transitions of the anthracene unit, in addition to the typical phthalocyanine absorption features. Notably, the Q-band appeared prominently at 690 nm. A relatively broader band was also observed, attributed to $n \rightarrow \pi^*$ transition between sulfur atoms and the porphyrazine core. Upon $Pd^{2+}$ addition (4.0 equiv) to the probe, both the Q-band and the $n \rightarrow \pi^*$ transition band displayed substantial intensity reductions, underlining the probe's responsive behavior toward $Pd^{2+}$ detection. Besides, the Q-band broadened and red-shifted because of the aggregates formation. Similar spectral changes have been observed for $Ag^+$ and $Hg^{2+}$ ions. This probe exhibited relatively low emission quantum yeild ($\Phi_{em}$ = 0.005) in THF solution. The reason behind the weak luminescence could be ascribed to the PeT process existing between anthracenyl and the porphyrazine moieties. Moreover, the quenching of the porphyrazine core

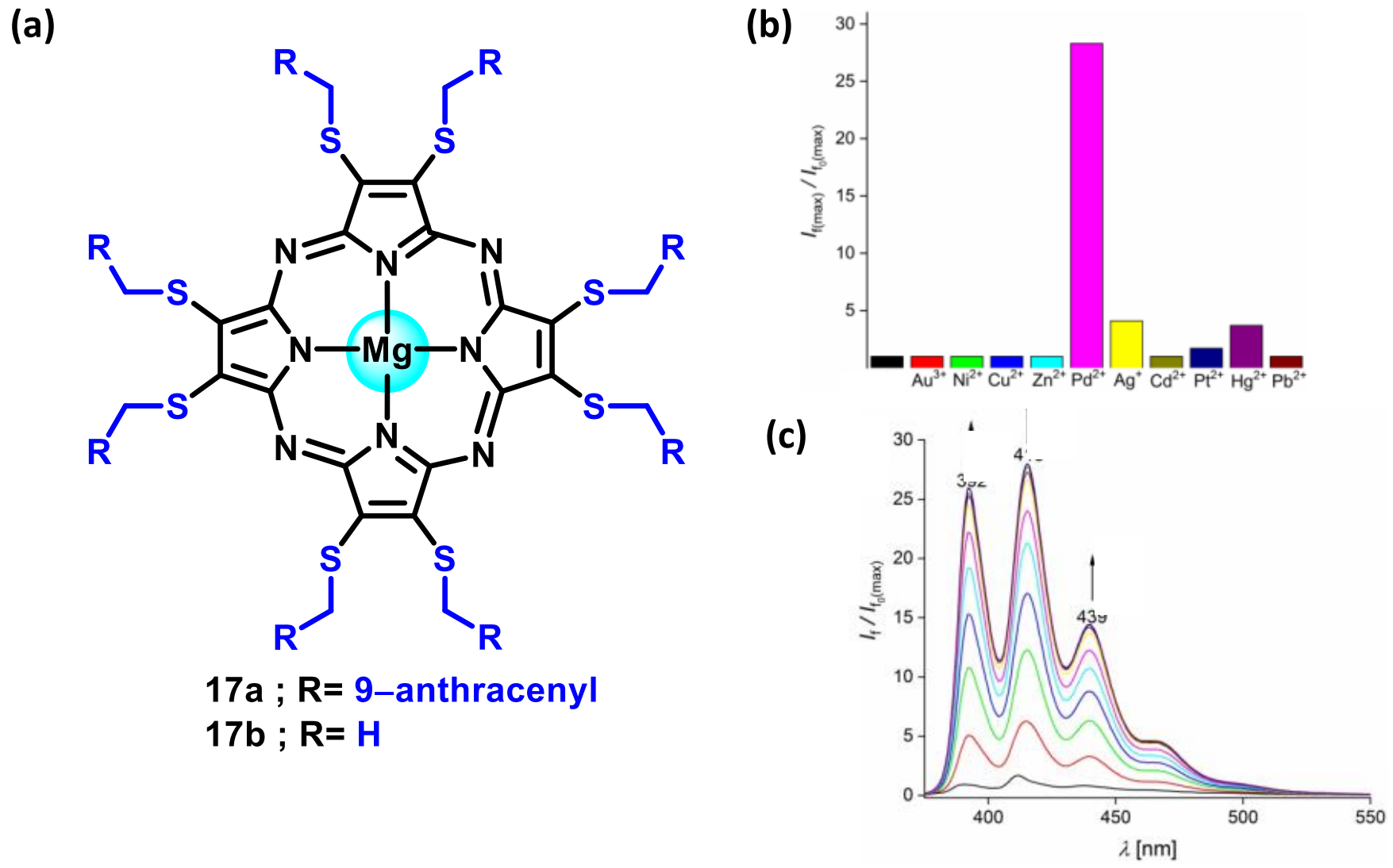

**Fig. 13** (a) Chemical drawings of $Mg^{2+}$-based phthalocyanines **17a** and **17b**; (b) luminescence intensity changes in THF solution upon addition of different metal ions and (c) luminescence enhancement in **17a** upon $Pd^{2+}$ addition in THF solution ($\lambda_{ex}$= 365 nm). Adapted with permission from ref [41], Copyright, Wiley.

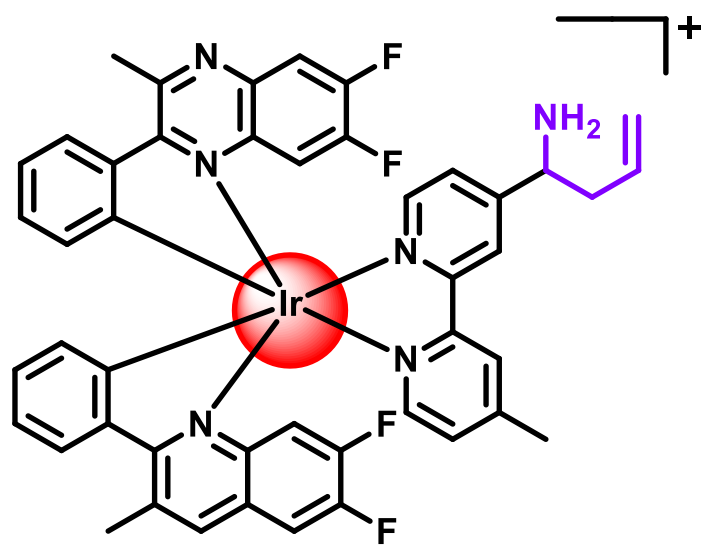

**18a**

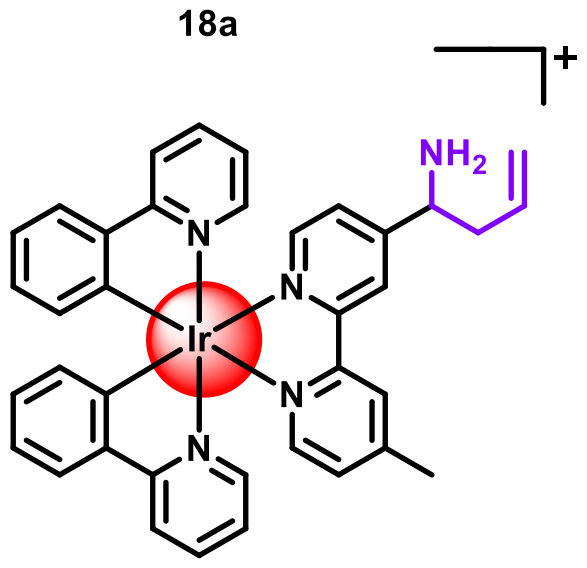

**18b**

**Fig 14.** Chemical drawings of $Ir^{3+}$-based probe **18a** and **18b**.

corresponds to the ET from the excited anthracene to the porphyrazine core. A remarkable luminescence enhancement in **17a** could be noticed when $Pd^{2+}$ was added up to 8.0 equiv (Fig. 13(c)) in THF solution. The ideal size of the S-N-S meso pocket acted as a key factor contributing to the luminescence enhancement, with probe **17a** achieving an impressively low detection limit of 0.70 µg/L.

In year 2023, Liu *et* al.[155] synthesized two NIR-active $Ir^{3+}$ based probes **18a** and **18b** incorporating amino-allyl groups for selective detection of $Pd^0$ and $Pd^{2+}$ ions (Fig. 14). These probes exhibited LC $\pi\longrightarrow\pi^*$ transitions in 250-320 nm range, with MLCT absorptions appearing between 320-400 nm. Upon excitation at 260 nm, **18a** and **18b** emitted at 685 nm and 618 nm, respectively. The luminescence lifetimes of **18a** and **18b** were reported to be 314.8 ns and 74.7 ns, respectively, with the corresponding quantum efficiencies of 6.41% and 1.12%. Though both the probes demonstrated high selectivity for Pd-based ions, detailed studies have been focussed on **18a** because of its NIR sensitive feature. The sensing studies were performed in $CH_3CN$-HEPES mixture. The addition of $Pd^0$ to **18a** resulted in a rapid luminescence quenching at 685 nm within 5 minutes, with a detection limit of 0.50 µM. The interference studies were also conducted against amino acids (such as serine, histidine, asparagine, methionine, homocysteine and cysteine) as well as various metal ions (such as, $Li^+$, $K^+$, $Ag^+$, $Na^+$, $Ba^{2+}$, $Ca^{2+}$, $Ni^{2+}$, $Fe^{2+}$, $Mn^{2+}$, $Ba^{2+}$, $Co^{2+}$, $Zn^{2+}$, $Cu^{2+}$, $Ba^{2+}$, $Cd^{2+}$, $Cr^{3+}$, $Al^{3+}$, $Fe^{3+}$ and $Au^{3+}$). Minor interferences were observed in case of $Cr^{3+}$ and $Fe^{3+}$ due to their weak coordination ability and with $Au^{3+}$ due to its moderated affinity for allyl group. However, the quenching efficiency was maximum for $Pd^0$ ions. This probe was capable of detecting $Pd^0$ in different Xi'an Moat water samples. In addition, the live-cell imaging studies utilizing probe **18a** demonstrated its ability to selectively target mitochondria. The probe was non-toxic to HeLa cells, with an $IC_{50}$ value of 19.65 µM within 6 hours. Bright red luminescence remarkably quenched upon addition of varied concentrations of $Pd^0$ and $Pd^{2+}$ ions. Co-localization experiments with Mito-Tracker Green, a mitochondrial dye, confirmed mitochondrial targeting. Photostability evaluations under 405 nm laser excitation showed a significant luminescence retention for probe **18a** even after 420 seconds, in contrast to the rapid fading of Mito-Tracker Green, evidencing its superior photostability.

## 5. Metal-organic frameworks (MOFs) based probes

Over the past two decades, an extensive range of MOFs have been utilized across diverse research areas including heterogeneous catalysis,[156–160] drug delivery,[161–163] gas adsorption and separation,[164,165] magnetic materials,[166,167] optical devices,[168,169] to name but a few. This advancement can be attributed to their intrinsic structural versatility and tunable chemical as well as photophysical properties.[170,171] MOFs are typically hybrid materials constructed by coordinating metal ions or metal clusters with organic ligands (acting as linkers), leading to the formation of one-dimensional (1D), two-dimensional (2D), or three-dimensional (3D) MOFs structures.[172–174] The careful selection of specific metal ions and organic linkers is pivotal in the development of desired MOFs.

Unlike many other materials, MOFs exhibit distinctive structural features, including open porous architectures, large surface areas, exceptional porosity, and tuneable pore surfaces, which combinedly contribute to their versatility and functionality.[175,176] Amongst these attributes, the use of MOFs as optical sensing materials (either luminescent or colorimetric), to detect small cations, anions, and neutral structural diversity and adjustability, and their chemical or photophysical properties can easily be tuned by changing either the metal centre or the organic ligand's frame employed to construct the probe.[177] Such materials having large specific surface area readily concentrate the target analyte which eventually results into an increased detection sensitivity; indeed, due to their high porosity, MOFs offer some pre-concentration effect recurrently leading to an upswing in molecules, stands out undeniably as one of the most imperative applications. Luminescent MOFs-based probes display probe's sensitivity for guest analytes.[49,178] Furthermore, the flexible porosity of MOFs defines the distance from the target analyte to the host MOF ensuring an appropriate interaction between the two. Also, high crystallinity of MOFs (whenever available) may impart proper structure-activity relationships (SARs) required to modify the MOFs as per the target application.

**Table 1**. Metal complexes based luminescent/colorimetric probes for $Pd^{2+}$ ion.

| Probe | $\lambda_{abs}$ (nm) | $\lambda_{em}$ (nm) | Target Ions | Quenching/ Enhancement | LoD (µM) | Binding constant ($K_b$, $M^{-1}$) | Solvent | [Ref.] |
|---|---|---|---|---|---|---|---|---|
| **$Pd^{2+}$ detection in organic medium** | | | | | | | | |
| **1a-1d** | 672-688 | – | **$Pd^{2+}$**, $Ag^{+}$ | Colorimetric | – | – | DMF/THF | [119] |
| **2a-2c** | 674-686 | – | **$Pd^{2+}$**, $Ag^{+}$ | Colorimetric | - | – | $CH_3OH$/THF | [120] |
| **3a-3b** | 688, 675 | – | **$Pd^{2+}$**, $Ag^{+}$ | Colorimetric | – | – | $CH_3OH$/THF | [121] |
| **4a-4c** | 695-710 | – | **$Pd^{2+}$**, $Ag^{+}$ | Colorimetric | – | – | $CH_3OH$/THF | [122] |
| **5a-5e** | 693-755 | – | **$Pd^{2+}$**, $Ag^{+}$ | Colorimetric | – | – | THF | [123] |
| **6a-6c** | 624-638 | – | **$Pd^{2+}$** | Colorimetric | – | -- | $CH_3OH$/THF | [124] |
| **9a-9d** | 693-720 | – | **$Pd^{2+}$**, $Ag^{+}$ | Colorimetric | – | – | THF | [125] |
| **10a-10b** | 707-708 | – | **$Pd^{2+}$**, $Ag^{+}$ | Colorimetric | – | – | THF | [126] |
| **11a-11b** | 704-706 | 722-723 | **$Pd^{2+}$**, $Ag^{+}$ | Colorimetric | – | – | THF | [127] |
| **12a-12b** | 704-733 | 728-754 | **$Pd^{2+}$**, $Ag^{+}$ | Colorimetric | – | – | THF | [127] |
| **13** | 1705 | 710 | **$Pd^{2+}$** | Quenching | – | – | THF | [137] |
| **15a-15c** | 686-704 | 704-721 | **$Pd^{2+}$** | Quenching | – | – | $CH_3OH$/THF | [154] |
| **16a-16c** | 701-706 | 722 | **$Pd^{2+}$**, $Ag^{+}$ | Quenching | – | $Ag^{+} = 1.4 \times 10^{8}$ **$Pd^{2+} = 3.4 \times 10^{7}$** | DMSO | [42] |
| **17a** | 690 | 415 | **$Pd^{2+}$** | Enhancement | 0.70 | – | THF | [41] |
| **$Pd^{2+}$ detection in semi-aqueous medium** | | | | | | | | |
| **7** | 356 | 615 | **$Pd^{2+}$**, $F^{-}$, | Enhancement | $80 \times 10^{-3}$ | | $CH_3CN$: pH=10 buffer | [43] |
| **14** | 562 | 558 | **$Pd^{2+}$** | Enhancement | $8.46 \times 10^{-3}$ | $4 \times 10^{6}$ | THF/ Water | [153] |
| **18a** | 320–420 | 685 | **$Pd^{2+}$** | Quenching | 0/50 | | $CH_3CN$/HEPES | [155] |
| **18b** | 320–420 | 618 | **$Pd^{2+}$** | Quenching | | | $CH_3CN$/HEPES | [155] |
| **$Pd^{2+}$ detection in water** | | | | | | | | |
| **8** | 428 | 670 | **$Pd^{2+}$** | Quenching | 1.54 | $3.2 \times 10^{3}$ | Water | [145] |

Based upon these peculiar advantageous features, a diverse panel of luminescent MOFs have been developed to detect metal ions, anions, small gases, organic molecules etc.[46,49,179,180]

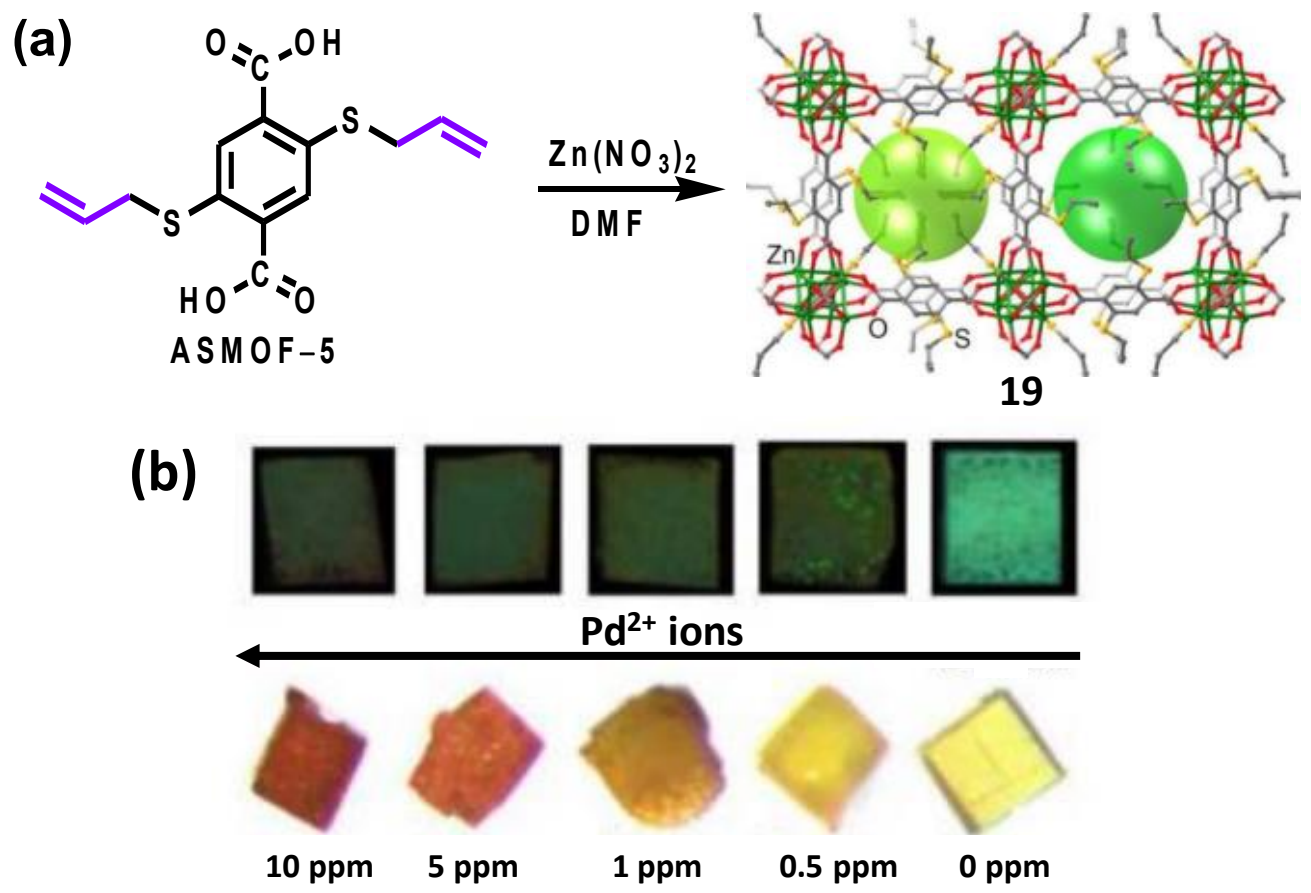

**Fig 15.** (a) Schematic for synthesis of probe **19** and (b) solid-state luminescent and colorimetric responses of **19** for $Pd^{2+}$ at various concentrations (0-10.0 ppm) in $CH_2Cl_2$ and $CH_3CN$, respectively ($\lambda_{ex}$= 365 nm). Adapted with permission from ref [181]. Copyright 2013, American Chemical Society.

while the MOFs substantially acting as colorimetric sensors are relatively less explored.

The following section specifically highlights the use of MOFs as optical sensing probes for detecting $Pd^{2+}$ ions. As per the reports available, the major research works on $Pd^{2+}$-responsive MOFs and coordinate metal polymers have centred around $Zn^{2+}$ and $Cd^{2+}$ metallic cores.[181–188] However, recent advancements have also introduced lanthanides-based MOFs acting as a compelling alternative for $Pd^{2+}$ sensing applications (Table 2).[189]

In year 2013, He *et* al.[181] demonstrated the solid-state colorimetric detection of $Pd^{2+}$ using $Zn^{2+}$-MOF **19** functionalized with alkene and thioether groups (Fig. 15(a)). The S-conjugated aromatic groups in **19** featured strong photoluminescence properties, typically required for the sensing purpose. In addition, the alkene moiety, having intrinsic $\pi$-donor and $\pi$-acceptor behaviors, binds precious and noble metals more preferably, thus enabling the fast uptake of soft metal ions. $Pd^{2+}$ is chemically soft acid and interacts strongly with S-containing functionalities (e.g., thioethers, thiols, etc.). Immersion of **19** crystals in $CH_3CN$ solution of $Pd^{2+}$ triggered a sharp color change of the crystals from pale yellow to orange brick-red in a couple of minutes, suggesting a strong interaction between solid MOF and $Pd^{2+}$ ions. This color change was found highly dependent on the varied $Pd^{2+}$ concentrations in solution. For instance, at 0.5 ppm $Pd^{2+}$, the yellow color of the crystals slightly faded and then turned to orange tinge color. Upon gradual increase in $Pd^{2+}$ concentration (upto 10.0 ppm), the crystals appeared much darker (Fig. 15(b)). Initially, other competing soft metal ions (e.g., $Pt^{2+}$, $Au^{+}$, $Ag^{+}$, and $Ru^{3+}$) failed to induce any color change; however, at their elevated concentration (*ca.* 1000 ppm), the MOF's crystals turned brown-grey in color.

In 2015, Konar's group[182] synthesized another $Zn^{2+}$-based luminescent MOF **20** that demonstrated the exceptional selectivity for $Pd^{2+}$ and picric acid. The alkene (-CH=CH-) groups present along the pore walls as well as inside the pores were primarily responsible for $Pd^{2+}$ recognition through $\pi \longrightarrow d$ transition between $Pd^{2+}$ and the alkene group. The luminescence spectrum of **20** in DMF displayed an intense emission at 415 nm coupled with a weak shoulder near 393 nm ($\lambda_{ex}$ = 340 nm). This shoulder has been ascribed to ligand-to-ligand-charge-transfer transition (LLCT) occurring between coordinated and non-coordinated alkene moieties. A rapid and substantial luminescence quenching (*ca*. 90%) was observed at 415 nm upon progressive addition of $Pd^{2+}$ to the probe's solution in DMF. Moreover, the solution visibly changed color from colorless to grey (obvious to the naked eye). UV-Vis spectroscopy was further employed to study the sensing ability of **20** toward $Pd^{2+}$ ions. A red-shifted absorption from 340 nm to 362 nm was observed with the emergence of a new band at 270 nm upon $Pd^{2+}$ addition to the probe's solution. The resultant spectral changes have been attributed to a strong alkene-$Pd^{2+}$ interaction. The inclusion of $Pd^{2+}$ caused a negligible change in the emission lifetime of probe from $5.11 \times 10^{-10}$ s to $5.13 \times 10^{-10}$ s, suggesting a static quenching mechanism. The $K_{SV}$ value of $3.6 \times 10^{4}$ $M^{-1}$ indicated the high efficiency of the probe for $Pd^{2+}$ which was corroborated by selectivity experiments in the presence of other competing species. The detection limit was computed as 0.03 ppm with 1:2 binding stoichiometry for **20-**$Pd^{2+}$ adduct.

Selective detection of metal ions in aqueous-phase using MOFs-based probes is often challenging, most likely due to the underlying solvation effect in such systems. To develop a MOF functioning in the aqueous media for practical applications, it is crucial to focus on the stability of MOF's framework in moist conditions or in pure water.[49,190] In year 2017, Parmar *et* al.[183] synthesized two MOFs bearing $Zn^{2+}$ (**21a**) and $Cd^{2+}$ (**21b**) metal cores to detect $Pd^{2+}$ in aqueous solutions. Indeed, this study marked the first reported use of luminescent MOFs for $Pd^{2+}$ recognition in aqueous medium. Probes **21a** and **21b** were isostructural and comprised of 4-pyridyl carboxaldehyde *iso*-nicotinoylhydrazone together with bipyridyl-derived Schiff base and 2-amino-terephthalic acid as the organic linkers. The emission spectra of **21a** and **21b** were measured by dispersing 3.0 mg of probes in 3.0 mL deionized water. The emission bands for **21a** and **21b** were measured at 428 nm and 431 nm, respectively, when excited at 340 nm. The sensing ability of **21a** and **21b** has been investigated for diverse panel of metal ions e.g., $Na^{+}$, $K^{+}$, $Mg^{2+}$, $Ba^{2+}$, $Ca^{2+}$, $Zn^{2+}$, $Cd^{2+}$, $Mn^{2+}$, $Ni^{2+}$, $Hg^{2+}$, $Co^{2+}$, $Cr^{3+}$, $Cu^{2+}$, $Pb^{2+}$, $Fe^{3+}$ and $Pd^{2+}$ in the aqueous solution (metal ion conc. was *ca.* $10^{-2}$ M).

Only a slight decrease in the emission intensity of probes could be noticed when exposed to $Ni^{2+}$, $Hg^{2+}$, $Co^{2+}$, $Cr^{3+}$ and $Cu^{2+}$ ions. On the other hand, the luminescence fully quenched for $Fe^{3+}$ and $Pd^{2+}$ which could be observed by the naked eye under UV light illumination ($\lambda_{max}$ 360 nm). Rest of the cations did not display any changes in the luminescent intensity of probes. The emission lifetimes (~15.0 ns) of **21a-21b** were found unchanged after the addition of $Fe^{3+}$ and $Pd^{2+}$ ions. The LoD values for $Pd^{2+}$ were reported to be 0.20 $\mu$M for **21a** and 0.10 $\mu$M for **21b**. The interaction between the target metal ion and accessible amino- and/or amide- group(s) in the organic linker was primarily responsible for the sensing behavior of these MOFs. Such interactions interfere with the energy/electron transfer processes in probes which could be further supported *via* ICP-MS and FT-IR studies. For practical applications, the filter-paper strip experiments have been performed in which a paper strip was coated with **21a** or **21b** and tested against a pool of cations. The sensing ability of **21a** and **21b** was also evaluated for various anions, and notably, the luminescence intensity of the probes completely quenched in the presence of metal oxo anions such as $CrO_4^{2-}$ and $Cr_2O_7^{2-}$ ions.

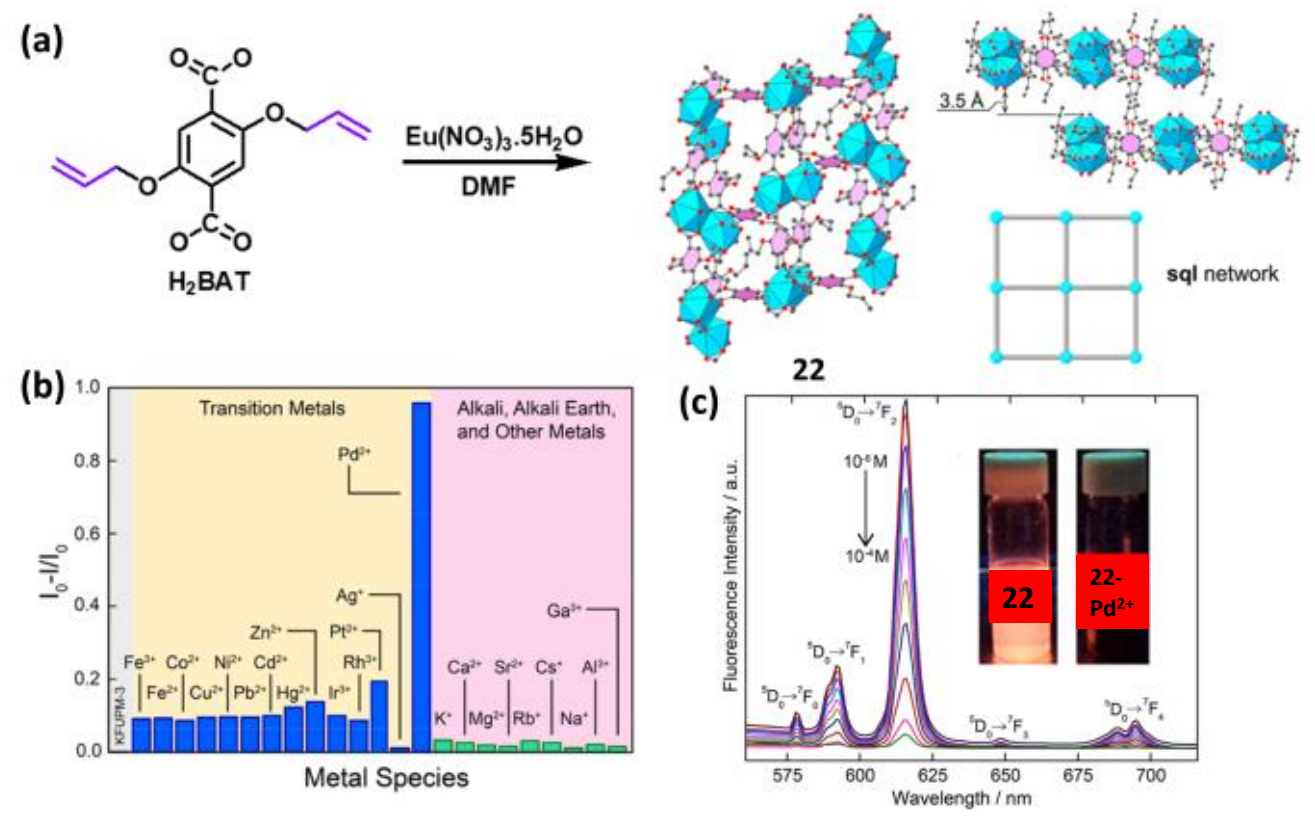

**Fig 16.** (a) Synthetic route to prepare probe **22**; (b) relative luminescence intensity of **22** in presence of different metal ions and (c) emission spectra of probe **22** upon successive addition of $Pd^{2+}$ ions in water ($10^{-6}$ M to $10^{-4}$ M) ($\lambda_{ex}$= 336 nm, pH = 7.3). Adapted with permission from ref [189]. Copyright 2019, American Chemical Society.

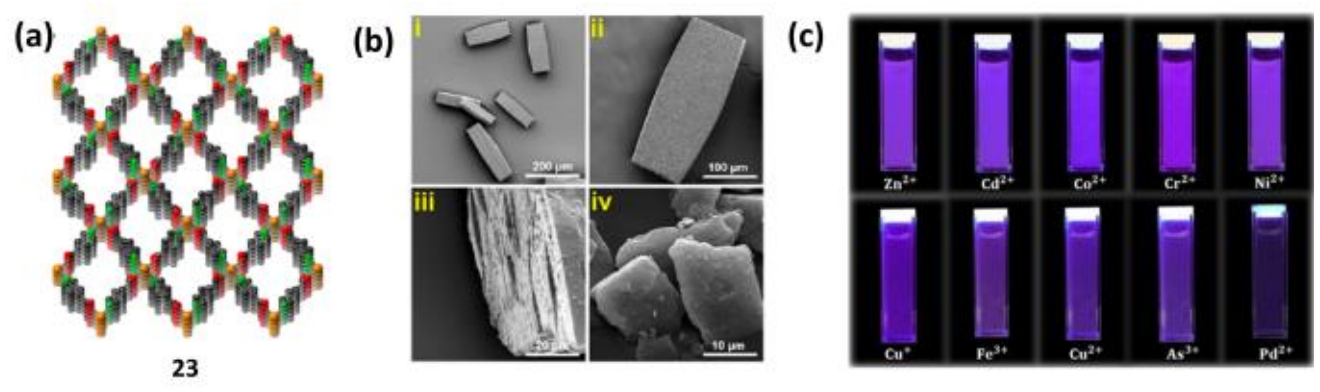

**Fig 17.** (a) Multi-layer 1D channel view of **23**; (b) SEM images of (i) bulk crystals of **23**, (ii) single crystal of **23**, (iii) 2D multilayer side-view morphology of **23** and (iv) exfoliated morphology of **23-NS** and (c) Color changes in **23** after the addition of assorted metal ions in $CH_3CN$ ($\lambda_{ex}$= 285 nm, pH = 4.0). Adapted with permission from ref [191]. Copyright 2019, American Chemical Society.

Another aqueous-phase detection of $Pd^{2+}$ was reported by Helal and co-workers[189] in year 2019, using a luminescent MOF **22**. This probe comprised of $Eu^{3+}$ core with 2,5-bis(allyloxy)terephthalic acid ($H_2BAT$) as the organic linker (acting as 'antenna' for $Eu^{3+}$) (Fig. 16(a)). UV-Vis spectrum of **22** revealed two $\pi\rightarrow\pi*$ transitions at 252 nm and 336 nm, attributed to the benzene ring and allyloxy groups. Upon $Pd^{2+}$ addition, a new band emerged at 276 nm suggesting an interaction between $Pd^{2+}$ and the allyloxy group in **22**. No significant changes could be realized upon the addition of $Na^+$, $K^+$, $Ca^{2+}$, $Mg^{2+}$, $Sr^{2+}$, $Rb^+$, $Cs^+$, $Fe^{3+}$, $Fe^{2+}$, $Co^{2+}$, $Cu^{2+}$, $Ni^{2+}$, $Zn^{2+}$, $Cd^{2+}$, $Ag^+$, $Al^{3+}$, $Ga^{3+}$, $Ir^{3+}$, $Rh^{3+}$ and $Pt^{2+}$ ions, suggesting the highly selective nature of **22** for $Pd^{2+}$ ion (Fig. 16(b)). A 1:2 binding stoichiometry was determined between the probe and $Pd^{2+}$ using Job's plot analysis. The emission spectrum of **22** exhibited characteristic $Eu^{3+}$ emissions at 578 nm, 592 nm, 616 nm, 648 nm and 695 nm due to the corresponding ${}^5D_0\longrightarrow{}^7F_0$, ${}^5D_0\longrightarrow{}^7F_1$, ${}^5D_0\longrightarrow{}^7F_2$, ${}^5D_0\longrightarrow{}^7F_3$ and ${}^5D_0\longrightarrow{}^7F_4$ transitions. The $Pd^{2+}$ addition led to decline the emission band of **22** that quenched completely at 2:1 stoichiometry of $Pd^{2+}$-**22** (Fig. 16(c)). The authors suggest that the $Pd^{2+}$ ions bind with the allyloxy groups of the ligand and interfere with the energy transfer mechanism from 'antenna' to $Eu^{3+}$ ion, ultimately resulting into the luminescence quenching response. Probe **22** was functional in the pH range 6.0-10.0, and selectively detected $Pd^{2+}$ even in presence of other competing species. The Stern-Volmer constant was reported to be 7.8 × $10^4$ $M^{-1}$ with a very low detection limit (i.e., 44 ppb).

In year 2019, Hmadeh and group[191] synthesized a $Cu^{2+}$ and isonicotinic acid (INA) based MOF **23** using solvothermal method, and fabricated the layered structure to 2D nanosheets **23-NS** by top-down liquid ultrasonication-exfoliation method (Fig. 17(a)). Probe **23** exhibited its emission maximum near 409 nm ($\lambda_{ex}$ 285 nm) that could be ascribed to the LC $\pi\rightarrow\pi*$ or $n\rightarrow\pi*$ transition. The luminescence sensing ability of **23-NS** was tested for various metal ions (with 2.5 ppm concentration) such as $Fe^{3+}$, $Fe^{2+}$, $Al^{3+}$, $As^{3+}$, $Mg^{2+}$, $Co^{2+}$, $Cd^{2+}$, $Ni^{2+}$, $Cu^{2+}$, $Cu^+$, $Mn^{2+}$, $Hg^{2+}$, $Zn^{2+}$, $Pd^{2+}$, $Cr^{3+}$, $Ca^{2+}$ and $Ag^+$, where solely $Pd^{2+}$ exhibited 90% luminescence quenching in the dispersed solution of **23-NS** in acetone. The interference from other analytes was negligible, confirming a highly selective nature of the probe. The luminescence titration experiments showed the quenching response at a very low concentration of $Pd^{2+}$ (i.e., 0.02 ppm), reaching saturation at 0.8 ppm. The resulting changes were obvious under the UV light illumination as the original pale violet color of the probe disappears in presence of $Pd^{2+}$ ions. The above-mentioned experiment was also performed on **23** to check the influence of exfoliation on the sensing of the cation. Probe **23** exhibited a lesser quenching response in comparison to **23-NS**. High selectivity of 2D nanosheets **23-NS** for $Pd^{2+}$ has been attributed to the availability of more exposed binding sites for $Pd^{2+}$ after exfoliation. The pH effect studies demonstrated a remarkable increment in the luminescence of **23-NS** as the pH of the medium was gradually lowered from 10.0 to 4.0. At pH 4.0, the luminescence was completely quenched in the presence of 0.57 ppm $Pd^{2+}$ ions. The $K_{sv}$ value was computed as 1.6 × $10^4$ $M^{-1}$ implying considerable luminescence quenching and a detection limit of 0.02 ppm. The luminescence quenching response occurs likely due to a strong interaction between the unsaturated alkene group of the INA ligand and $Pd^{2+}$, facilitated *via* π → d transition.

Nandi *et* al.[192] constructed $Al^{3+}$-based MOF **24** integrated with an unsaturated vinyl group to detect $Pd^{2+}$ *via* $\pi\rightarrow d$ interactions. UV-Vis spectrum of **24** showed an absorption band at 305 nm that disappeared upon addition of 2.0 mM aqueous solution of $Pd^{2+}$. The resulting spectral changes have been ascribed to $\pi\rightarrow d$ interactions between MOF's functionality and $Pd^{2+}$ ion. The emission spectra of **24** have been monitored at 365 nm in the presence of a diverse range of the metal ions. Addition of $Pd^{2+}$ in an aqueous suspension of **24** exhibited the luminescence quenching upto *ca.* 90% within 30 minutes. Nominal emission changes were noticed for other cations e.g., $Ag^{+}$, $Cr^{3+}$, $Cu^{2+}$, $Cd^{2+}$, $Co^{2+}$, $Eu^{3+}$, $Fe^{2+}$, $Mg^{2+}$, $Pb^{2+}$, $Mn^{2+}$, $Ni^{2+}$, $Ru^{3+}$ and $Zn^{2+}$, except a slight quenching in case of $Fe^{3+}$ (~20%) and $Hg^{2+}$ (~22%). The luminescence lifetime of this probe (i.e., 2.69 ns) was unaltered by the presence of $Pd^{2+}$ ion. The pH effect studies clearly indicated that the probe could function at a wider pH range from 2.0 to 12.0. Probe **24** showed a relatively lower LoD value i.e., 19.50 ppb, and selectively recognised $Pd^{2+}$ even in presence of other interfering and competing analytes. The Stern-Volmer plot was linear at the low concentration and non-linear at a higher concentration, suggesting the existence of both static as well as the dynamic quenching mechanisms. Notably, probe **24** was also capable of detecting Pd in different oxidation states exhibiting variable quenching efficiency (70-90%).

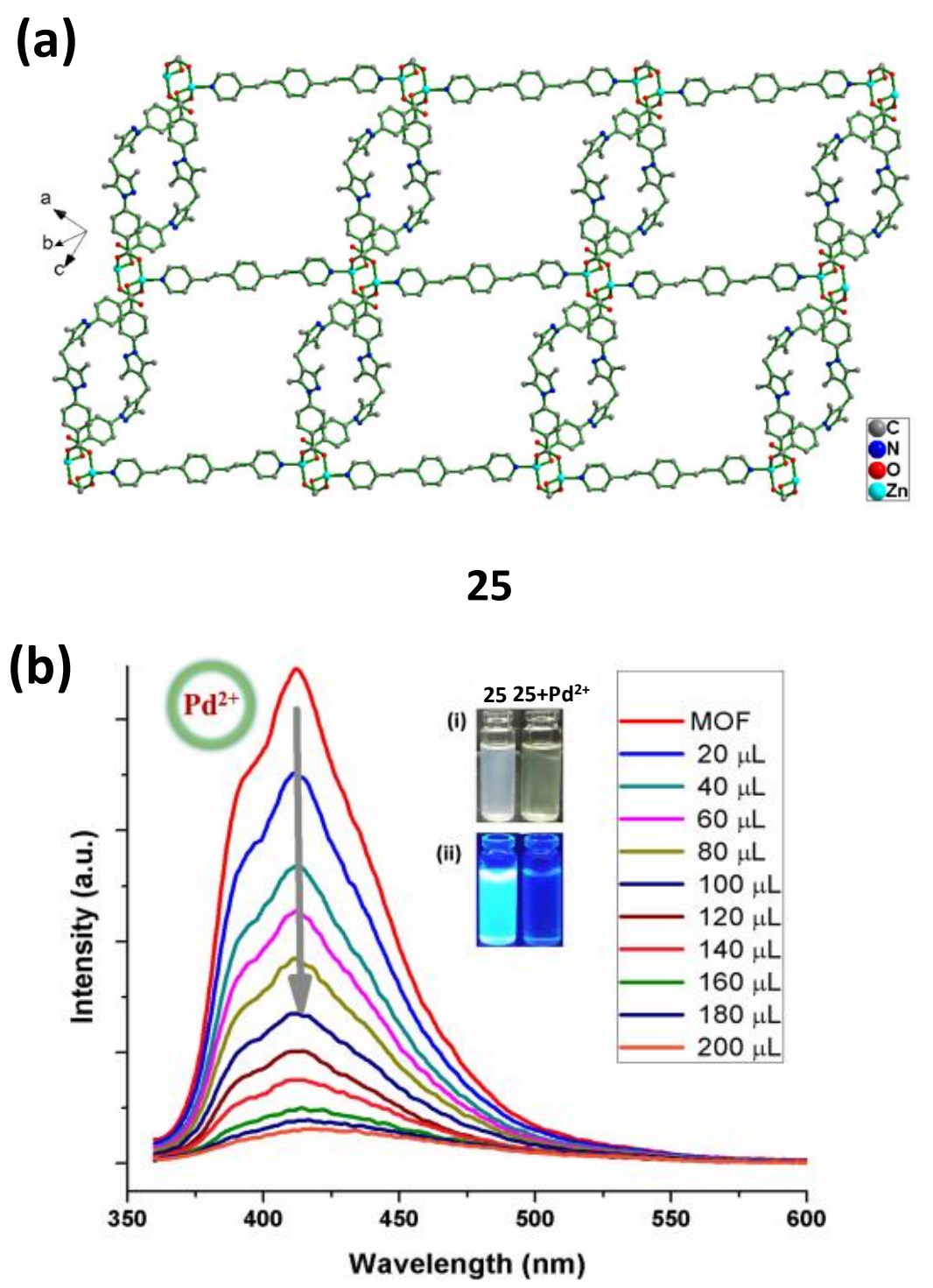

**Fig 18.** (a) 2D-sheet structure of **25** exhibiting the coordination pattern and (b) spectral changes observed in the luminescence intensity of **25** at 414 nm upon $Pd^{2+}$ (1.0 mM) addition in DMF solution. Inset: obvious color changes in probe **25** with $Pd^{2+}$ upon exposure to (i) visible and (ii) UV light in DMF solution ($\lambda_{ex}$= 270 nm). Adapted with permission from ref [185]. Copyright 2019, Wiley.

In year 2019, our group[185] reported a $Zn^{2+}$ and *bis*-pyrazole based 2D MOF **25** for selective luminescent recognition of $Pd^{2+}$ ions. This was the first report on such a highly luminescent porous MOF bearing open alkenes and nitrogen sites along the walls favourable for target metal ions (Fig. 18(a)). UV-Vis spectrum of **25** exhibited an absorption peak at 350 nm while its emission maximum appeared at 414 nm upon excitation at 270 nm in DMF solution. Owing to the favourable electronic environment of open pores, the sensing ability of **25** was investigated for various metal ions such as $Co^{2+}$, $Ni^{2+}$, $Cu^{2+}$, $Mn^{2+}$, $Mg^{2+}$, $Zn^{2+}$, $Cd^{2+}$, $Hg^{2+}$, $Ag^{+}$, $Au^{+}$, $Pt^{2+}$, $Rh^{3+}$, $Ir^{3+}$ and $Pd^{2+}$. Nitrate salts of the above-mentioned metals were added to the DMF suspension of **25**, however, only $Pd^{2+}$ could induce almost complete quenching of the luminescence (*ca.* 94%). A slight quenching was also noticed in case of strongly competing $Pt^{2+}$, $Ag^{+}$, $Au^{+}$, $Rh^{3+}$ and $Ir^{3+}$ ions. In the solid-state, **25** acted both as colorimetric and luminescent sensor to recognize $Pd^{2+}$, and the changes were obvious to the naked eye (Fig. 18(b)). Notably, upon $Pd^{2+}$ addition, the absorption spectrum of **25** displayed *ca.* 22 nm red-shift in 350 nm band with the emergence of a new peak near 270 nm. The selective detection of $Pd^{2+}$ by **25** has been attributed to the accessible open unsaturated >C=C< groups along the walls of the MOF that display strong binding with soft $Pd^{2+}$ ions. Job's plot analyses indicated a 1:2 stoichiometry between $Pd^{2+}$ and **25**. Additionally, the reusability of **25** was investigated over three cycles by restoring its emission with the addition of EDTA ions (*ca.* 0.1 M) to **25**-$Pd^{2+}$ adduct.

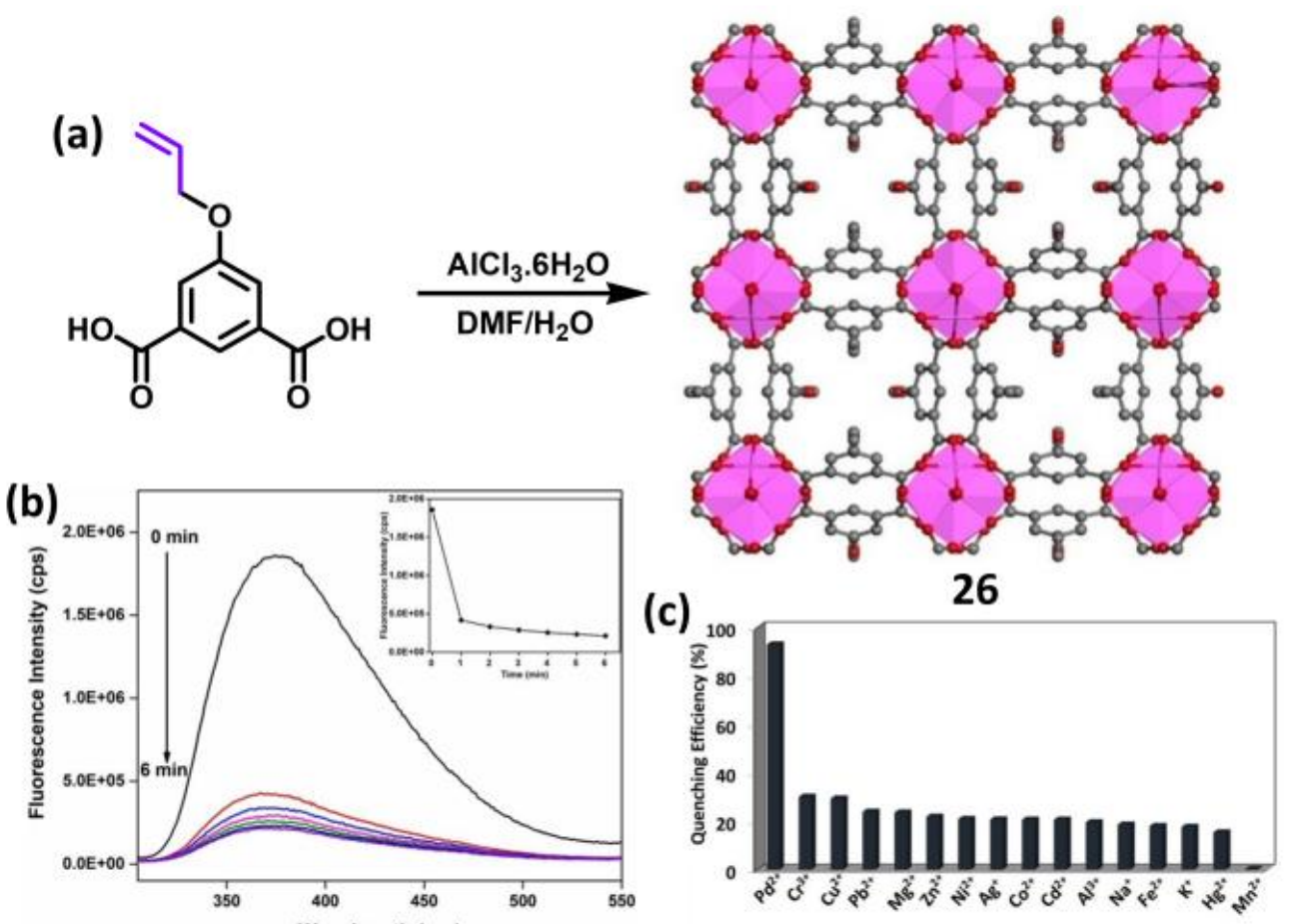

**Fig 19.** (a) Synthesis of $Al^{3+}$-MOF based probe **26**; (b) Time-dependent luminescent changes in **26** after the addition of $Pd^{2+}$ ions and (c) Luminescence quenching effect of metal ions employed with probe **26** in water ($\lambda_{ex}$= 290 nm). Adapted with permission from ref [193] Copyright 2021, Wiley.

**Table 2**. Metal-organic frameworks (MOFs) based luminescent/colorimetric probes for $Pd^{2+}$ ion.

| Probe | $\lambda_{abs}$ (nm) | $\lambda_{em}$ (nm) | Target Ions | Quenching/ Enhancement | LoD (µM) | Binding constant ($K_b$, $M^{-1}$) | Solvent | [Ref.] |
|---|---|---|---|---|---|---|---|---|
| **$Pd^{2+}$ detection in organic medium** | | | | | | | | |
| **19** | | **524** | **$Pd^{2+}$** | Colorimetric | **500** | - | $CH_3CN$ | [181] |
| **20** | 423 | 415,393 | PA, **$Pd^{2+}$** | Quenching | $0.03 \times 10^3$ | - | DMF | [182] |
| **23** | | 409 | **$Pd^{2+}$** | Quenching | $0.02 \times 10^3$ | -- | $CH_3CN$ | [191] |
| **25** | | 414 | **$Pd^{2+}$** | Quenching | $4.7 \times 10^{-1}$ | | DMF | [185] |
| **31** | | 620 | **$Pd^{2+}$** | Quenching | $62.3 \times 10^{-3}$ | | DMSO | [198] |
| **$Pd^{2+}$ detection in water** | | | | | | | | |
| **21a** | 365 | 428 | $CrO_4^{2-}$, $Cr_2O_7^{2-}$, $Fe^{3+}$, **$Pd^{2+}$** | Quenching | $CrO_4^{2-} = 0.25$<br>$Cr_2O_7^{2-} = 0.43$<br>$Fe^{3+} = 3.76$<br>**$Pd^{2+} = 0.20$** | - | Water | [183] |
| **21b** | | 431 | $CrO_4^{2-}$, $Cr_2O_7^{2-}$, $Fe^{3+}$, **$Pd^{2+}$** | Quenching | $CrO_4^{2-} = 0.18$<br>$Cr_2O_7^{2-} = 0.19$<br>$Fe^{3+} = 1.77$<br>**$Pd^{2+} = 0.10$** | - | Water | [183] |
| **22** | 252, 336 | 616 | **$Pd^{2+}$** | Quenching | 44 | | Water | [189] |
| **24** | 292 | 365 | **$Pd^{2+}$**, $H_2S$ | Quenching | $Pd^{2+} = 110 \times 10^{-3}$,<br>$H_2S = 1.65 \times 10^{-3}$ | | Water | [192] |
| **26** | | | **$Pd^{2+}$** | Quenching | $1.48 \times 10^{-1}$ | | Water | [193] |
| **27** | | 425 | **$Pd^{2+}$** | Quenching | $0.011 \times 10^3$ | | Water | [187] |
| **28** | | 370 | **$Pd^{2+}$** | Quenching | 0.102 | | Water | [196] |
| **29** | | 341,408 | **$Pd^{2+}$**, PA | Quenching | $0.52 \times 10^{-3}$ | | Water | [184] |
| **30** | | 445 | **$Pd^{2+}$** | Quenching | $0.0072 \times 10^3$ | | Water | [186] |
| **32** | 315 | 412 | **$Pd^{2+}$** | Quenching | 0.08 | | Water | [188] |

In year 2021, $Al^{3+}$-MOF (probe **26**) incorporating 5-(allyloxy)-isophthalic acid as an organic linker was presented by Ghosh *et al.*[193] for luminescence-based detection of $Pd^{2+}$ in aqueous solution (Fig. 19(a)). The emission maximum of **26** remarkably declined at 378 nm upon addition of $Pd^{2+}$ (10.0 mM) to its dispersed aqueous solution. The limit of detection was calculated as 26.24 ppb (or 0.148 µM) with a $K_{sv}$ value of $1.05 \times 10^5$ $M^{-1}$. Authors claimed that the $K_{sv}$ value shown by **26** was the highest amongst the $Pd^{2+}$-responsive MOFs, reported so far. The inclusion of $Pd^{2+}$ resulted in significant changes to UV-Vis spectrum of MOF **26**. The absorption near 306 nm showed an increase in intensity along with a 6 nm red-shift, while a new band emerged at 420 nm. These spectral changes have been attributed to the strong interaction between $Pd^{2+}$ and the allyloxy functionalities present in the MOF structure.

It has been observed that a MOF equipped with the olefinic double bonds can respond to the light-induced photochemical

[2+2] cycloaddition reactions, relying on a particular alignment of the double bonds in its solid-state. Such MOFs experience a sudden change in their volume *via* photochemical [2+2] cycloaddition, and are known to display ‘photo-salient effect’.[187,194] In an elegant report (in 2022), Bera *et* al.[187] constructed a $Zn^{2+}$-based 2D MOF **27** bearing 4-(1-naphthylvinyl)pyridine and succinic acid as the organic linkers. This probe was capable of recognising $Pd^{2+}$ ions in aqueous media and exhibited ‘photo-salient effect’ under UV light illumination (Figs. 20(a-b)) Similar to a kernel-popping on hot plate, the crystals of MOF **27** blowed up and exploded upon UV light exposure. Almost same but a slow effect could be seen in the sunlight as well.

Probe **27** emitted at 425 nm in aqueous solution when excited at 340 nm; however, the addition of $Pd^{2+}$ (1.0 mM) induced a significant quenching in its luminescence intensity. A quenching effect upto *ca.* 98% could be observed with only 1.06 ppm concentration of $Pd^{2+}$. A non-linear behavior of Stern-Volmer plot ($K_{sv} = 3.8 \times 10^5$ L/mol) at a higher $Pd^{2+}$ concentration suggested the existence of both static as well as dynamic quenching mechanisms. The LoD value was computed to be 0.011 ppm which is much lower than the permissible $Pd^{2+}$ value in drinking water as set by WHO.[195] Other metal ions such as $Cr^{3+}$, $Al^{3+}$, $Fe^{3+}$, $Cu^{2+}$, $Hg^{2+}$, $Cd^{2+}$, $Co^{2+}$, $Mn^{2+}$, $Pb^{2+}$, $Ni^{2+}$, $Zn^{2+}$, $Na^{+}$, $Ca^{2+}$, $K^{+}$ and $Ba^{2+}$ did not display any luminescence changes in **26** and failed to interfere with the detection of $Pd^{2+}$ by **27** (Fig. 20(c)). The selective sensing mechanism was attributed to the strong binding of $\pi$-electrons rich olefinic and aromatic moieties with $Pd^{2+}$ ions. The metal ion interaction was further established with the help of NMR analyses by titrating probe **27** with $Pd^{2+}$ ions. The luminescence lifetime of **27** enhanced from 0.3368 ns to 0.8656 ns upon $Pd^{2+}$ inclusion.

In a recent report (in 2023), Biswas and coworkers[196] reported another $Al^{3+}$-based MOF **28** functionalized with propynyl-oxy group for ultrafast and highly sensitive/selective detection of $Pd^{2+}$ in water. This probe exhibited an emission band spanning from 310-550 nm upon excitation at 290 nm. The emission band was significantly quenched upon $Pd^{2+}$ addition due to its binding with the alkyne group. The probe showed an impressive response time of just 30 seconds with a LoD value of 102.2 nM. Interestingly, the addition of $Pd^{2+}$ resulted in negligible luminescence lifetime changes, suggesting that the recognition was achieved by the formation of ground-state complex. Other metal ions including $Pt^{2+}$, $Hg^{2+}$, $Cu^{2+}$, $Pb^{2+}$, $Cd^{2+}$, $Mg^{2+}$, $Co^{2+}$, $Cr^{3+}$, $Ag^{+}$, $Zn^{2+}$, $Ni^{2+}$, $Mn^{2+}$, $Zn^{2+}$, $Na^{+}$, $Fe^{2+}$, $Al^{3+}$, $K^{+}$, $Eu^{2+}$, $Cs^{+}$, $In^{3+}$, $Gd^{3+}$, and $Pr^{3+}$ caused no interference in the selective detection of $Pd^{2+}$ ion by **28**. Furthermore, this probe was found to be recyclable; simple washing restored its fluorescence even after five cycles, and EDX and PXRD studies evidenced no palladium retention in the probe, indicating no chemical reaction occurred. Instead, the $Pd^{2+}$ binding was attributed to the alkyne group interaction. UV-Vis analysis further supported such an analyte interaction, as the original absorption of **28** at 302 nm increased significantly along with a red-shift of 8 nm upon $Pd^{2+}$ addition to **28**. More importantly, this probe was efficient in detecting $Pd^{2+}$ ions in real water sources, including distilled water, river water, tap water, and lake water. For practical applications, low-cost filter paper strips were developed by depositing probe **28** onto their surface. Upon drying, the paper strips exhibited blue fluorescence, which disappeared upon $Pd^{2+}$ addition when observed under a fluorescence lamp.

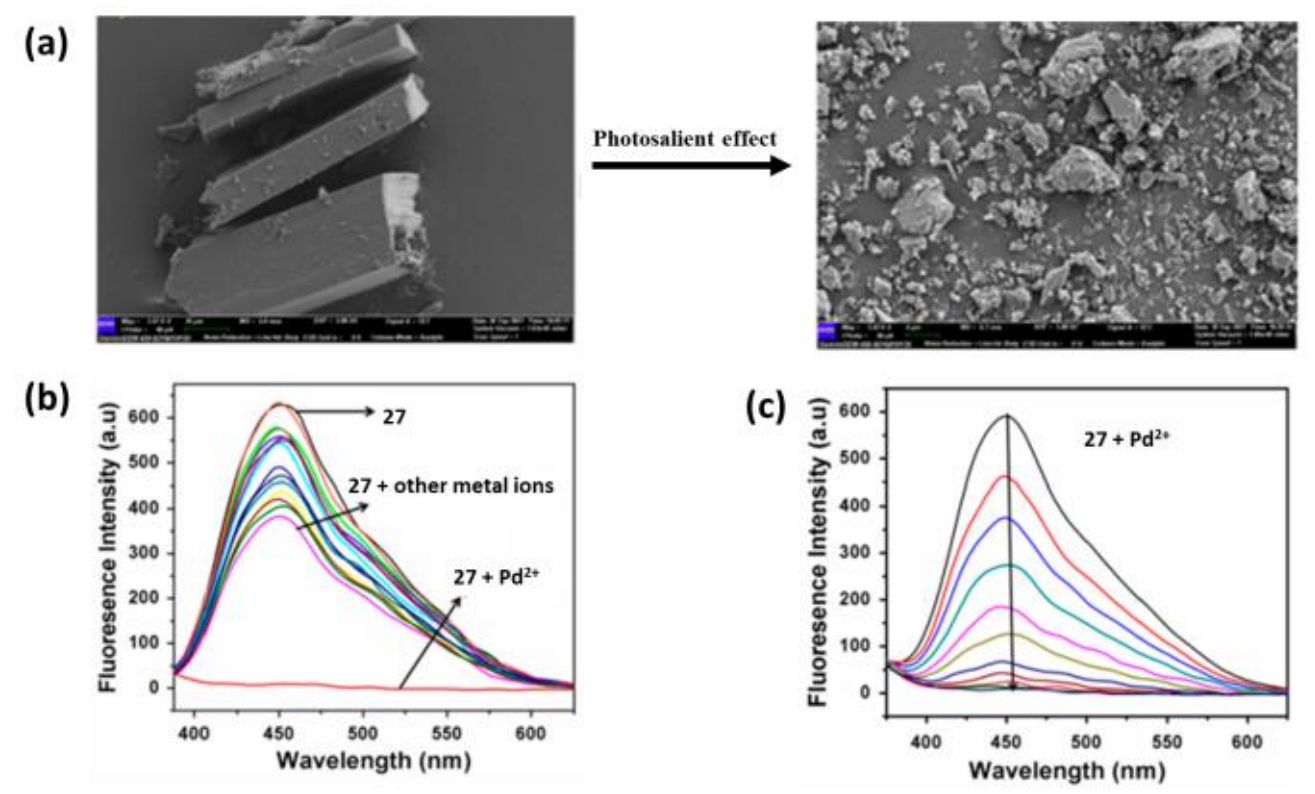

**Fig 20.** (a) FE-SEM images of probe **27** (a) before and after UV light exposure; (b) Luminescence quenching effect of various metal ions employed with probe **27** and (c) Luminescence quenching in **27** at different concentrations of $Pd^{2+}$ ions in water ($\lambda_{ex}$= 340 nm). Adapted with permission from ref [187]. Copyright 2022, American Chemical Society.

## 6. Metal-polymer coordinates as $Pd^{2+}$-responsive probes

In comparison to traditional metal complexes and MOFs as luminescent sensing probes, the metal-based 1D coordination polymers are under-explored for selective sensing of analytes.[184,197] During the recent past, such polymers have emerged as viable sensing platforms for heavy metal ions. In year 2020, Ashafaq *et* al.[184] presented a $Zn^{2+}$-based luminescent 1D coordination polymer **29** containing 2,3-pyridinecarboxylic acid as ligand for dual sensing of $Pd^{2+}$ and picric acid in aqueous solutions. Excitation of **29** at 274 nm resulted into two emission bands at 341 nm and 408 nm, with the luminescence lifetime value as $4.79 \times 10^{-8}$ s. The sensing behavior of **29** was tested for different metal ions in water. The emission spectra of **29** exhibited noticeable luminescence quenching (*ca*. 82%) upon the addition of 11.0 ppb of $Pd^{2+}$ ions (1.0 mM). An obvious color change from colorless to greyish yellow could also be noticed after $Pd^{2+}$ addition. The limit of detection has been computed as 0.52 ppb which is far below than the permissible value set by WHO. The Stern-Volmer constant ($K_{SV} = 0.466 \times 10^2$ $M^{-1}$) was much less than the values reported for various afore-mentioned MOFs, and the resultant non-linear plot suggested both static as well as dynamic quenching sensing mechanisms. No changes in the luminescence spectra could be observed upon the addition of other cations e.g., $Mn^{2+}$, $Co^{2+}$, $Ni^{2+}$, $Cu^{2+}$, $Cd^{2+}$, $Hg^{2+}$, $Pb^{2+}$ and $Pt^{2+}$ ions. The observed spectral changes in **29** have been attributed to the high affinity of $Pd^{2+}$ for unsaturated alkene groups in the polymer’s framework.

In the subsequent year (2021), Mir and group[186] synthesized a $Cd^{2+}$-based 1D polymer **30** having 4-(1-naphthylvinyl)-pyridine and 5-sulfosalicylic acid as ligands. This 1D probe underwent a [2+2] cycloaddition reaction under sunlight exposure to afford a cyclobutene-based 2D polymer **30'** in the solid-state (crystal-to-crystal transformation). Interestingly, the reaction was found reversible as the original 1D polymer could be recovered just by heating of **30'** (Fig. 21(a)). Structural changes between **30** and **30'** remarkably influenced their metal ion sensing properties. Probe **30** was reported to display a superior luminescence response toward $Pd^{2+}$ in comparison to **30'**. Upon excitation at 340 nm, probes **30** and **30'** showed their emission maxima at 445 nm and 448 nm, respectively, in aqueous media. The sensing ability of these polymers was evaluated against various metal ions such as $Na^+$, $Al^{3+}$, $Cu^{2+}$, $Hg^{2+}$, $Cd^{2+}$, $Co^{2+}$, $Mn^{2+}$, $Fe^{3+}$, $Pb^{2+}$, $K^+$, $Ba^{2+}$, $Ni^{2+}$, $Zn^{2+}$, $Pd^{2+}$, $Ca^{2+}$ and $Cr^{3+}$ (Fig. 21(b)). The luminescence intensity of **30** in water showed significant quenching (*ca*. 97 %) even at a low concentration (i.e., 0.08 ppm) of $Pd^{2+}$ (Fig. 21(c)). No remarkable changes have been noticed for other assorted metal ions. The $K_{SV}$ values were computed to be $2.55 \times 10^5$ L/mol and $2.53 \times 10^5$ L/mol with LoD values as 0.0089 ppm and 0.0072 ppm for **30** and **30'**, respectively. The reason for the luminescence quenching lied behind the strong interaction of $Pd^{2+}$ with the unsaturated aromatic ring in the polymers. The increase in the luminescence lifetime of **30** upon $Pd^{2+}$ addition also suggested the strong interaction of receptor with $Pd^{2+}$. $^1H$ NMR titration analyses supported the $Pd^{2+}$ binding mechanism. Upon $Pd^{2+}$ addition, few resonances in **30** showed broadening and shifting (0.1-0.2 ppm). These shifts were slightly more pronounced in case of **30'**, indicating a relatively stronger $Pd^{2+}$ interaction in the 2D polymer framework as compared to **30** with 1D framework.

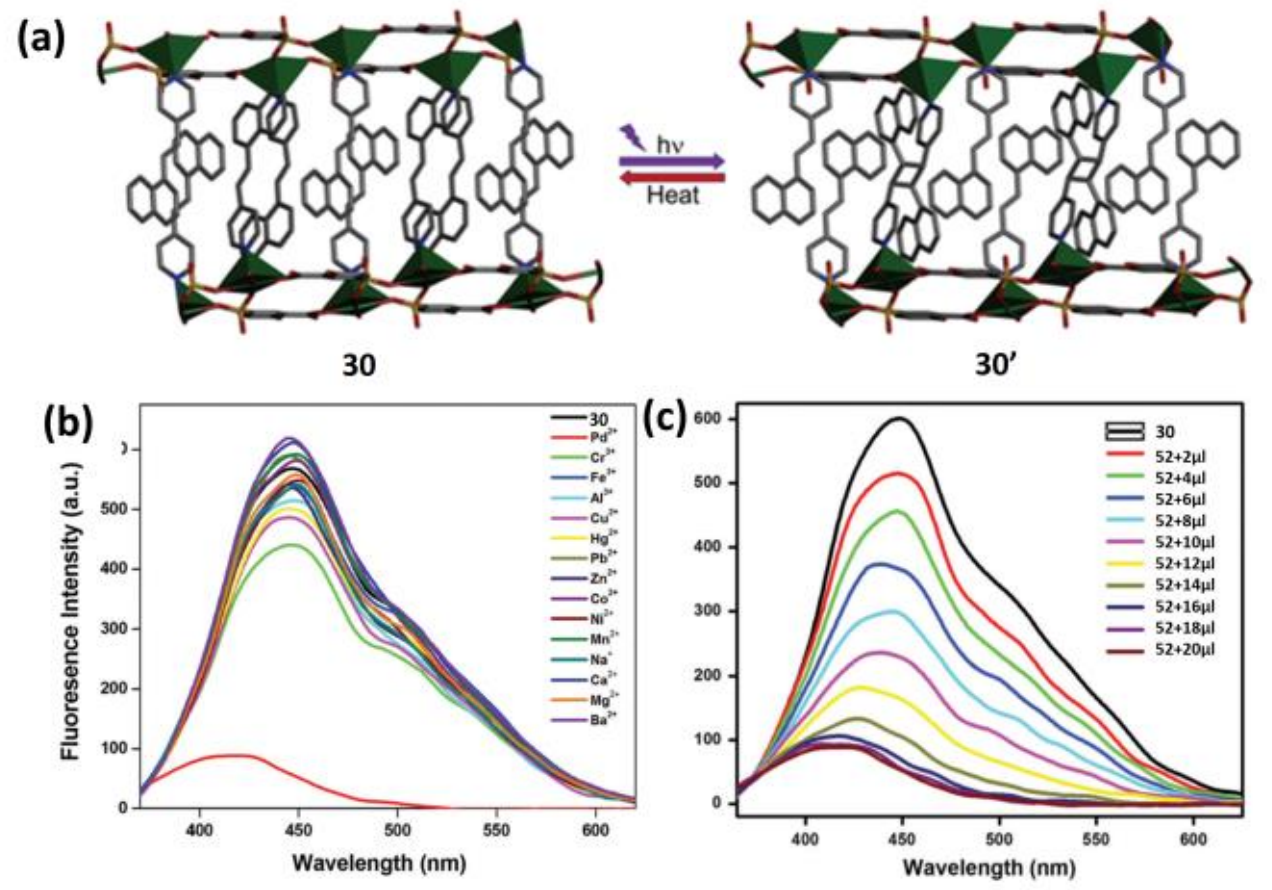

**Fig 21.** (a) Single-crystal to single-crystal structural transformation between **30** and **30'**; (b) Emission spectra of **30** upon addition of different metal ions and (c) Quenching of the emission band of **30** upon successive $Pd^{2+}$ addition in aqueous medium ($\lambda_{ex}$= 340 nm). Adapted with permission from ref [186]. Copyright 2021, The Royal Society of Chemistry.

In 2022, Lai *et* al.[198] demonstrated a $Zn^{2+}$-templated quadrangular-prismatic covalent cage **31** as a luminescent probe for $Pd^{2+}$ recognition. Probe **31** was constructed using 1,3-propanediammine and 5,5′,5′′,5′′′-(ethene-1,1,2,2-tetrayltetrakis(benzene-4,1-diyl))tetrapicolinaldehyde as organic linkers. In its solid-state, the probe **31** emitted at 572 nm ($\lambda_{ex}$ 446 nm) whereas in solution phase (in DMSO), the emission maximum shifted to 554 nm when excited at 393 nm. Notably, the luminescence quantum yield was also higher in the solid-state suggesting an aggregation-induced emission behavior in the probe. The addition of 10% water by volume in $2.0 \times 10^{-5}$ mol/L DMSO solution of **31** led to the luminescence enhancement upto 2.2-fold; however, further water addition resulted in the quenching response likely due to either proton or electron transfer process existing between chromophore and water. To test the metal ion sensing ability of **31**, various metal ions e.g., $Pd^{2+}$, $Fe^{3+}$, $Cr^{3+}$, $Fe^{2+}$, $Cd^{2+}$, $Mg^{2+}$, $Co^{2+}$, $Ba^{2+}$, $Ca^{2+}$, $Pb^{2+}$, $Pt^{2+}$, $Na^+$ and $Ni^{2+}$ were added to DMSO solution of **31** ($2.0 \times 10^{-4}$ mol/L). Though $Ni^{2+}$ addition caused a luminescence enhancement upto 2.74-fold, $Pd^{2+}$ induced almost complete luminescence quenching (*ca.* 99%) in probe **31**. The quenching effect has been attributed to the replacement of $Zn^{2+}$ by $Pd^{2+}$ resulting into the heavy atom effect. This displacement was thermodynamically favourable as this was accompanied with the release of the coordinated water molecules in the parent cage which ultimately increased the overall entropy of the system. This cage-shaped probe was highly selective for $Pd^{2+}$ with the estimated LoD value as 62.3 nM in DMSO solution.

Recently, Sinha and coworkers[188] reported another coordination polymer (**32**) for selective detection of $Pd^{2+}$ in aqueous media. This 2D polymer consisted of $Cd^{2+}$ coordinated to the 5-aminoisophthalate and 2,3,5,6,-tetrakis(2-pyridyl)pyrazine having free terminal -COOH groups. UV-Vis studies of **32** displayed an absorption band at 315 nm together with a shoulder near 363 nm. On the other hand, its emission band could be seen at 412 nm. The addition of various metal ions such as $Al^{3+}$, $Fe^{3+}$, $Cd^{2+}$, $Cr^{3+}$, $Mn^{2+}$, $Zn^{2+}$, $Co^{2+}$, $Ni^{2+}$, $Pb^{2+}$, $Cu^{2+}$, $Ba^{2+}$, $Hg^{2+}$, $Mg^{2+}$, $Ca^{2+}$, $Na^+$ and $K^+$ to **32** induced no changes in its emission intensity. However, the emission intensity completely quenched upon the addition of $Pd^{2+}$ ions, with no interference from the assorted ions in the selective detection of $Pd^{2+}$. Furthermore, the fluorescence lifetime of **32** increased from $4.51 \times 10^{-9}$ s to $6.28 \times 10^{-10}$ s upon $Pd^{2+}$ addition. The high selectivity as well as the structural stability of this coordination polymer toward $Pd^{2+}$ ions have been attributed to the free -COOH group and π-cloud interactions. The probe exhibited a low limit of detection of 0.08 μM with a Stern-Volmer constant of $7.2 \times 10^4$ $M^{-1}$.

## 7. Some notable challenges in metal-organic $Pd^{2+}$ receptors

### 7.1 Sensing approaches in metal-organic receptors

Most metal complexes based optical probes employ colorimetric methods for $Pd^{2+}$ detection, with phthalocyanine derivatives being the predominant choice. However, these derivatives are hindered by complex synthesis processes and low yields. Though some phthalocyanine-based probes (e.g., probes **13**, **15b**, **15c**, and **16b**) demonstrate fluorogenic properties, their reliance on 'turn-off' responses limits precision. Furthermore, their sensing performance has mainly been tested in organic solvents, restricting their practical application. Recent progress includes the introduction of a magnesium-based phthalocyanine probe (**17a**) by Schwarze, which exhibits a 'turn-on' response in THF solution. However, such advancements remain inadequate in aqueous environments, where utility is paramount. Additionally, the limited development of fluorescence-based probes is problematic, as these probes often exhibit binary 'turn-on' or 'turn-off' emission responses. Such responses are susceptible to environmental fluctuations, photobleaching, and concentration-dependent variations, potentially leading to false positives. Ratiometric probes, which display measurable changes in the intensity of multiple emission bands upon interaction with $Pd^{2+}$ ions, offer improved reliability and analytical precision. Similarly, in the case of MOFs, most reported probes for $Pd^{2+}$ detection utilize 'turn-off' mechanisms in water. Although these MOFs exhibit poor water solubility, their suspensions have shown potential for sensing applications.

### 7.2 Lack of systematic evaluation of interfering cations

A significant challenge in metal-organic $Pd^{2+}$ receptors lies in the incomplete and inconsistent evaluation of selectivity against competing metal ions. Soft cations like $Au^{+}$, $Ag^{+}$, and $Pt^{2+}$, which exhibit similar coordination preferences to $Pd^{2+}$, are often overlooked or insufficiently tested. Though soft donor groups favor $Pd^{2+}$ binding, they can inadvertently bind to these competing ions if not carefully optimized. Strategic modifications, such as introducing steric hindrance or secondary coordination features, are necessary to enhance specificity. Moreover, matrix composition including pH, ionic strength, and organic content can dramatically influence competitive binding, highlighting the need for matrix-specific interference studies. Environmental samples (e.g., seawater, wastewater) and biological matrices (e.g., serum, cytosol) present additional complexity, often not accounted for in current designs. Establishing standardized, comprehensive selectivity protocols will be essential to advance practical $Pd^{2+}$ sensing.

### 7.3 Addressing photostability challenges

Photostability is a critical factor in ensuring consistent and reliable performance of optical sensors. However, majority of the studies have inadequately investigated this property. Among the reported metal complexes, photostability assessments have only been performed for a limited number of probes (e.g., **11a**-**11b**, **12a**-**12b**, **15b**-**15c** and **18a**-**18b**). Similarly, while MOFs and coordination polymers often demonstrate chemical and thermal stability, their photostability has received minimal attention, highlighting a significant research gap.

### 7.4 Real-world application challenges

Despite advances in probe design, real-world deployment of $Pd^{2+}$ sensors remains limited. Most studies assess performance under idealized laboratory conditions, with few validating efficacies in complex matrices like soil, drinking water, seawater, or food. These environments introduce factors such as high salinity, heavy metal interference, and organic matter, all of which can impair probe performance. In biological contexts, although probes **14** and **18** have shown promise for $Pd^{2+}$ imaging in HeLa cells, expanding studies to additional cell lines like A549 and CHO is crucial to validate broader biological relevance. Further testing in biological fluids and under physiologically relevant conditions is needed to establish clinical utility.

To bridge laboratory success and field applications, future studies must address probe robustness, matrix tolerance, regeneration ability, and operational stability. Environmentally friendly, scalable sensor designs will be critical for broader adoption in environmental monitoring and biomedical diagnostics.

### 7.5 pH sensitivity

The pH sensitivity of metal-organic receptors plays a crucial role in their performance, yet this parameter remains significantly underexplored for many $Pd^{2+}$ sensors. Phthalocyanine-based probes, for example, are seldom evaluated across varying pH conditions; a notable exception is probe **15**, which displayed optimal fluorescence at neutral pH, although its $Pd^{2+}$ sensing capability was not examined, likely due to the susceptibility of phthalocyanine to protonation or deprotonation, which can disrupt metal coordination and optical properties. Coordination polymers and MOFs often suffer from structural degradation under extreme pH, though certain metal complexes have demonstrated exceptional stability. Notably, Ru-based probe **8** remained functional across a wide pH range (1.1-12.2), and probe **18a** could detect $Pd^{2+}$ under neutral and basic conditions.

Among MOFs, pH-dependent structural integrity was evident. For instance, probe **22** maintained its framework between pH 6-10, while PXRD revealed decomposition beyond this range. Probe **27** exhibited stable fluorescence across a broad pH span (2–12), but without $Pd^{2+}$ detection studies. In contrast, probe **23** effectively detected $Pd^{2+}$ at pH 4, 7, and 10, with complete fluorescence quenching observed at pH 4. These findings highlight the need for next-generation receptors that maintain both structural and functional integrity under extreme pH conditions ($pH < 4$ or $> 9$). Receptors with strong chelating groups like imidazole or thiolates and rigid, hydrophobic structures show better stability across varying pH levels. Strategies such as placing active sites in hydrophobic MOF cavities or adding polymer coatings can further improve durability. Future designs should focus on pH-resistant systems, especially for use in complex environments where pH changes are common.

### 7.6 Need for on-site and portable detection systems

On-site detection methods for $Pd^{2+}$ remain largely unexplored. Incorporating portable detection platforms, such as smartphone-based systems, could significantly enhance the practicality and accessibility of these sensors. In this regard, test paper strips (e.g., probes **21** and **28**) offer a promising solution due to their low cost, portability, disposability, and potential integration into future portable devices. Incorporating advanced computational tools, such as artificial intelligence (AI) and machine learning (ML), could revolutionize sensor design by optimizing parameters, improving detection accuracy, and broadening the scope of $Pd^{2+}$ sensing technologies.

## 8. Conclusions and future perspectives

This review highlights the design and performance of various metal-organic receptors, particularly metal complexes and MOFs, for the optical detection of $Pd^{2+}$ ions through luminescent and colorimetric approaches. Compared to small organic probes, metal-based systems exhibit superior sensitivity/selectivity, tunable optical properties, redox- and photo-stability, magnetic behavior, and enhanced solubility in aqueous environments. Detection strategies predominantly rely on binding interactions governed by Pearson's HSAB principle, with soft donor atoms playing a critical role in achieving high selectivity for $Pd^{2+}$ ions. Both luminescent and colorimetric methods have distinct strengths and limitations. Luminescent probes offer high sensitivity, real-time monitoring, and minimal background noise, but often require sophisticated instrumentation and may perform poorly under turbid conditions. Colorimetric probes, while simple and cost-effective for rapid, on-site testing, usually suffer from lower sensitivity and are prone to interference by environmental impurities. Thus, probe selection must consider the specific operational conditions, detection limits, and cost constraints.

Detailed studies of $Pd^{2+}$-responsive metal complexes underline the importance of optimizing receptor geometry, binding site configuration, and donor atom identity to enhance selectivity. $Pd^{2+}$, being a soft acid, prefers coordination with soft donor groups such as thiols, thioethers, sulfides, and phosphines, often adopting square planar geometries. Yet, designing probes with high tolerance to competing species like $Pt^{2+}$, $Ag^{+}$, and $Au^{+}$ and ensuring stability and solubility in real-world environments remains challenging.

In the case of MOF-based probes, the presence of >C=C< moiety often contributes to the preferential $Pd^{2+}$ binding, although the underlying mechanistic understanding is often incomplete. Despite impressive porosity and modularity, several MOFs frequently suffer from poor stability and solubility in aqueous media, limiting their practical deployment. Future efforts thus should focus on developing water-stable and water-soluble MOF architectures specifically tailored for $Pd^{2+}$ detection.

Importantly, combining materials, such as integrating metal complexes within MOF scaffolds or hybridizing MOFs with carbon-based composites (e.g., graphene oxide, carbon nanotubes, etc.), have emerged as a powerful strategy to overcome intrinsic material limitations.[199,200] These hybrid systems usually offer synergistic effects, improved stability, better optical response, increased surface area for analyte binding, and promoting selective interactions. Metal-complex/MOF composites, for instance, benefit from the structural robustness of MOFs and the tuneable binding environments of metal complexes, while carbon-based platforms contribute excellent conductivity, photostability, as well as mechanical durability.[201,202] Such integrative approaches are poised to lead the next generation of high-performance $Pd^{2+}$ sensors with superior analytical performance in complex and variable matrices. Additionally, emerging technologies like mixed-matrix membranes (MMMs) incorporating MOFs, and embedding metal nanoparticles into porous frameworks, further widen the scope of application by creating flexible, portable, and regenerable sensing platforms.[203–205] The use of near-infrared (NIR)-active upconverting materials also shows promise by minimizing background interference, since most competing species do not absorb NIR light.[49,206–208]

Moving forward, it is essential to establish detailed and rational Structure-Activity Relationships (SARs) for receptor design, systematically correlating receptor architecture, donor group nature, coordination environment, and sensing mechanisms with detection performance. Such an approach, coupled with computational modeling and high-throughput screening, will greatly accelerate the optimization of sensors for $Pd^{2+}$ and other critical analytes.

*In fine*, the authors hope that the diverse collection of $Pd^{2+}$ sensors discussed herein provides a valuable foundation for future innovation. By combining the strengths of different material classes, rigorously addressing real-world sensing challenges, and implementing systematic design strategies, it is possible to develop the next generation of highly robust, sensitive, and selective $Pd^{2+}$ sensors for broad environmental, biomedical, and industrial applications.

## Symbols/Abbreviations:

| | |
|---|---|
| Pd | palladium |
| MOFs | metal-organic frameworks |
| MPcs | metal-phthalocyanines |
| LoD | limit of detection |
| $K_b$ | binding or association constant |
| $K_{SV}$ | Stern-Volmer constant |
| $\Phi_{em}$ | emission quantum yield |
| nM | nanomolar |
| mM | millimolar |
| μM | micromolar |
| ppm | parts per million |
| LC | ligand-centred |
| ICT | intramolecular charge transfer |
| MLCT | metal-to-ligand charge transfer |
| LMCT | ligand to-metal charge transfer |
| UV | ultraviolet |
| NIR | near-infrared |
| HOMO | highest occupied molecular orbital |
| LUMO | lowest unoccupied molecular orbital |
| PeT | photoinduced electron transfer |
| FE-SEM | field emission scanning electron microscopy |
| HR-TEM | high-resolution transmission electron microscopy |
| SAR | structure-activity relationship |

## Author contributions

We strongly encourage authors to include author contributions and recommend using CRediT for standardised contribution descriptions. Please refer to our general author guidelines for more information about authorship.

## Conflicts of interest

The authors declare that they have no known competing financial interests or personal relationships that could have appeared to influence the work reported in this paper.

## Data availability

A data availability statement (DAS) is required to be submitted alongside all articles. Please read our full guidance on data availability statements for more details and examples of suitable statements you can use.

## Acknowledgements

SL, TG, and SK express their gratitude to the Central Instrumentation Centre (CIC) at UPES Dehradun for providing access to its instrumentation facilities. They also thank UPES for supporting their research through SEED (UPES/R&D-SoAE/08042024/2) and SHODH (UPES/R&D-SoAE/10042024/8) funding programs. Sudhanshu acknowledges the PhD fellowship support received from UPES Dehradun. Additionally, SK extends thanks to the Department of Science & Technology, New Delhi, India, for the INSPIRE Research Grant (IFA15/CH-213)